\newif\ifconfver
\confvertrue       
\ifconfver
\documentclass[10pt,twocolumn,twoside]{IEEEtran}
\else
\documentclass[11pt,draftcls,onecolumn]{IEEEtran}
\fi

\usepackage{makecell}
\usepackage{bbding}
\usepackage{color,graphicx}
\usepackage{float}
\usepackage{multirow,amsmath,epsfig,amsfonts,amssymb,psfig,graphics,psfrag,theorem,calc,url,bm,cite}
\usepackage{stfloats,hyperref}
\usepackage{algorithm,algorithmic}
\usepackage{diagbox} 
\usepackage{hhline}
\usepackage{subfigure}
\usepackage{pifont}
\usepackage{physics}
\usepackage{booktabs}
\usepackage{tikz}
\usepackage{multirow}
\usetikzlibrary{quantikz}
\usepackage{arydshln}
\usepackage{mathtools}

\def\blue{\color{black}}

\makeatletter
\def\multilimits@{\bgroup
	\Let@
	\restore@math@cr
	\default@tag
	\baselineskip\fontdimen10 \scriptfont\tw@
	\advance\baselineskip\fontdimen12 \scriptfont\tw@
	\lineskip\thr@@\fontdimen8 \scriptfont\thr@@
	\lineskiplimit\lineskip
	\vbox\bgroup\ialign\bgroup\hfil$\m@th\scriptstyle{##}$\hfil\crcr}
\def\Sb{_\multilimits@}
\def\endSb{\crcr\egroup\egroup\egroup}
\makeatother

	{\end{list}}

\newlength{\twidth}
\ifconfver
\else
\fi

\definecolor{orange}{RGB}{255,107,0}
\def\blue{\color{black}}

\theorembodyfont{\rmfamily}

{\begin{list}{}{
			\settowidth{\labelwidth}{\mbox{\textnormal{#1}}}%
			\setlength{\leftmargin}{\labelwidth+\labelsep}}}%
	{\end{list}}

\newcommand\bA{\ensuremath{{\bm A}}}
\newcommand\bB{\ensuremath{{\bm B}}}

\newcommand\bD{\ensuremath{{\bm D}}}
\newcommand\bE{\ensuremath{{\bm E}}}

\newcommand\bI{\ensuremath{{\bm I}}}

\newcommand\bP{\ensuremath{{\bm P}}}

\newcommand\bT{\ensuremath{{\bm T}}}
\newcommand\bU{\ensuremath{{\bm U}}}
\newcommand\bV{\ensuremath{{\bm V}}}

\newcommand\bX{\ensuremath{{\bm X}}}
\newcommand\bY{\ensuremath{{\bm Y}}}
\newcommand\bZ{\ensuremath{{\bm Z}}}

\newcommand\bv{\ensuremath{{\bm v}}}

\newcommand\bx{\ensuremath{{\bm x}}}

\definecolor{orange}{RGB}{255,107,0}
\def\blue{\color{black}}

\ifconfver

\author{Chia-Hsiang Lin,~\IEEEmembership{Senior Member,~IEEE}, Jian-Kai Huang,~\IEEEmembership{Student Member,~IEEE},
\\
Si-Sheng Young,~\IEEEmembership{Student Member,~IEEE}, and Wei-Cheng Zheng,~\IEEEmembership{Member,~IEEE}}

\title{Interpretable Landsat-to-Hyperspectral Dual Super-Resolution Without Large Matrix Inversion

\thanks{This study was supported by the Emerging Young Scholar Program (namely, the 2030 Cross-Generation Young Scholars Program) of National Science and Technology Council (NSTC), Taiwan, under Grant NSTC 114-2628-E-006-002.
We thank the National Center for Theoretical Sciences (NCTS) and the National Center for High-performance Computing (NCHC) for providing the computing resources.
\textit{(Corresponding Author: Chia-Hsiang Lin)}}
\thanks{C.-H. Lin is with the Department of Electrical Engineering, National Cheng Kung University (NCKU), Tainan 70101, Taiwan (R.O.C.) 
(e-mail: chiahsiang.steven.lin@gmail.com).}
\thanks{J.-K. Huang and S.-S. Young are with the Institute of Computer and Communication Engineering, Department of Electrical Engineering, National Cheng Kung University (NCKU), Tainan 70101, Taiwan (R.O.C.)
(e-mail: q38141509@gs.ncku.edu.tw, and q38121509@gs.ncku.edu.tw).}
\thanks{W.-C. Zheng is with the Agency of Rural Development and Soil and Water Conservation (ARDSWC), Ministry of Agriculture, Executive Yuan, Taiwan (R.O.C.)
(e-mail: bigbear44@mail.ardswc.gov.tw).}
}
\else

\fi

\begin{document}
	
	\bibliographystyle{IEEEtran}
	\maketitle
	\ifconfver \else \vspace{-0.5cm}\fi

\begin{abstract}
Direct acquisition of global hyperspectral images (HSIs) is infeasible given contemporary hardware facilities and limited resources, while global hyperspectral monitoring is critical for remote sensing applications.
%
%
A more economical approach is to \textit{interpretably} convert the global Landsat-8/9 multispectral images into NASA's AVIRIS-level HSIs.
This conversion involves both spatial super-resolution (SpaSR, 30-m to 15-m) and the highly ill-posed spectral super-resolution (SpeSR, 7-band to 172-band), referred to as the dual super-resolution (DualSR) as the duality between SpaSR and SpeSR has been derived very recently.
Existing SpeSR methods, mostly for reconstructing CAVE-level HSIs (31 visible bands only), are not applicable to AVIRIS-level task (172 visible and near/shortwave-infrared bands), motivating us to customize an interpretable alternating direction method of multipliers network (ADMM-Net) using Woodbury W-Lemma and spectral continuity prior.
Also, unlike conventional generative SpaSR, we employ the panchromatic sharpening strategy to recover physically grounded spatial details.
This strategy, however, induces very large matrix inversions (LMIs, with dimensionality proportional to the number of pixels) even after applying the W-Lemma.
We resolve the dilemma by trickily designing an LMI-free proximal gradient descent network (PGD-Net).
Consequently, a breakthrough in terms of complexity and performance is achieved by the proposed PGD-ADMM interpretable network (PAINT). 
Beyond state-of-the-art performance, PAINT significantly advances the Landsat classification performance (78.98\% accuracy and 76.06\% kappa) to a new level (92.16\% accuracy and 90.98\% kappa).
Source codes: \url{https://github.com/IHCLab/PAINT}
\end{abstract}


\begin{IEEEkeywords}
Landsat-8/9 satellite,
hyperspectral image,
identification,
spectral super-resolution,
panchromatic sharpening,
proximal gradient descent,
deep unfolding,
interpretable AI.
\end{IEEEkeywords}
	
	\ifconfver \else \vspace{-0.0cm}\fi
	
	\ifconfver \else \vspace{-0.5cm}\fi
	
	\ifconfver \else  \fi







\section{Introduction}\label{sec: introduction}
In recent years, hyperspectral images (HSIs) with wide spectral coverage and high spectral resolution (SpeR) have played a critical role in the remote sensing (RS) field \cite{MU2026TIP,liu2025cd,xu2026transgcf,wang2026amhf}.
For instance, NASA’s Airborne Visible/Infrared Imaging Spectrometer (AVIRIS), one of the most widely used hyperspectral systems \cite{brodrick2025aviris,CODE}, captures wavelengths from the visible to the shortwave infrared (SWIR) with a SpeR of 10 nm, thereby enabling numerous advanced Earth observation (EO) applications \cite{precision_agri, coleman2026identifying, mancoridis2025multi, sun2026hyperspectral}.
Owing to rich spectral information provided by hyperspectral data, global hyperspectral monitoring is highly critical for precise material and mineral mapping, as well as object identification and tracking.
Nevertheless, direct acquisitions of global HSIs remain practically infeasible due to current hardware restrictions and limited resources; for example, AVIRIS primarily operates over North America \cite{AVIRISdata}.
As a result, a more economical alternative is to algorithmically convert global multispectral imagery into its hyperspectral counterpart.
%
\begin{figure}[t]
\centering
\includegraphics[width=1\linewidth]{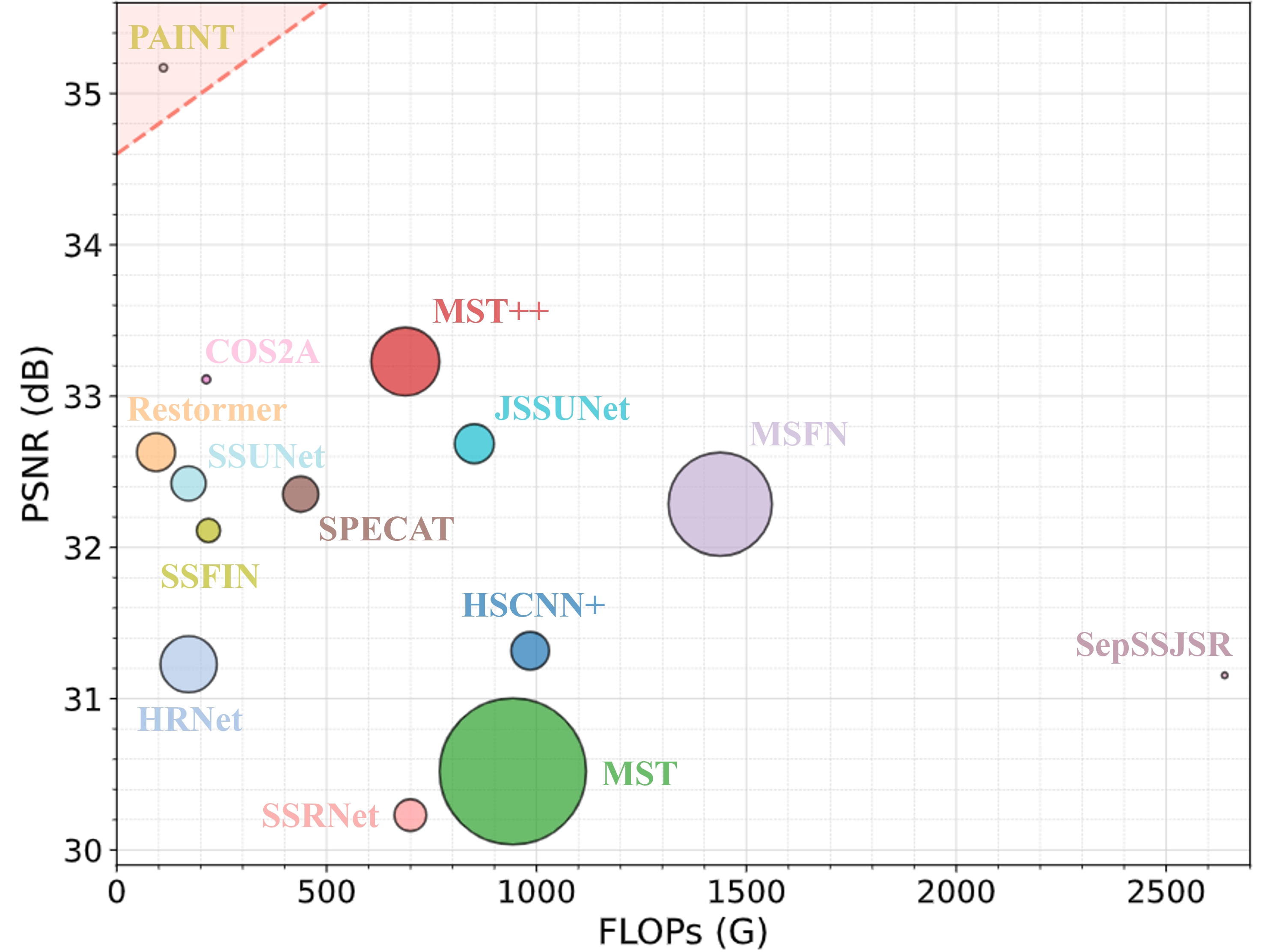}
\caption{Performance-Params-FLOPs comparisons. 
All methods perform DualSR from a 7-band Landsat-8/9 multispectral data to a 172-band AVIRIS HSI. 
The horizontal axis is computational complexity (measured in FLOPs), the vertical axis indicates performance in terms of PSNR, while the circle radius corresponds to the network parameters (memory cost). 
The FLOPs are computed based on a typical input size of 256 $\times$ 256 pixels.
We compare the proposed PAINT with 13 representative benchmark models, including SSFIN \cite{ma2022multi}, COS2A \cite{COS2A}, SPECAT \cite{SPECAT}, Restormer \cite{Restormer}, HRNet \cite{HRNet}, MST++ \cite{MST++}, SSRNet \cite{SSRNet}, MST \cite{MST}, HSCNN+ \cite{HSCNN}, MSFN \cite{MSFN}, JSSUNet \cite{JSSUNet}, SSUNet \cite{SSUNet}, and SepSSJSR \cite{SepSSJSR}.
}\label{fig:Landsat_FLOPs}
\end{figure}
\begin{table*}[t]
\centering
\caption{Comprehensive configurations of Landsat-8/Landsat-9 sensors.
The spectral channels ``CA'' and ``Pan'' denote ``coastal aerosol'' and ``panchromatic'', respectively. 
%
}
\label{tab:LANDSAT89}
\begin{tabular}{c|c|c|c|c|c|c|c|c|c|c|c}
\hline\hline

{Spectral Band} & {B1} & {B2} & {B3} & {B4} & {B5} & {B6} & {B7} & {B8} & {B9} & {B10} & {B11} 
\\
\hline

{Spectral Channel} & CA & Blue & Green & Red & NIR & SWIR-1 & SWIR-2 & Pan & Cirrus & TIRS-1 & TIRS-2 
\\
\hline\hline

{Landsat-8 Central Wavelength (nm)} & 440 & 480 & 560  & 655 & 865 & 1610  & 2200 & 590 & 1370 & 10895  & 12005 
\\
\hline

{Landsat-8 Bandwidth (nm)} & 20 & 60 & 60 &  30 & 30 & 80  & 180 & 180 & 20 &590  &1010
\\
\hline\hline

{Landsat-9 Central Wavelength (nm)} & 440 & 480 & 560  & 650 & 860 & 1605  & 2195 & 585 & 1370 & 10895  & 12005   
\\
\hline

{Landsat-9 Bandwidth (nm)} & 20 & 60 & 60 & 40 & 20 & 90  & 190 & 170 & 20 & 590  &1010
\\
\hline\hline

{Spatial Resolution (m)} &  30 & 30 & 30 & 30 & 30 &  30 & 30 & 15 & 30 & 100 & 100
\\
\hline\hline 

\end{tabular}
\end{table*}
\noindent
The key to achieving this impactful yet challenging RS application is NASA’s longest-running EO program, Landsat.

Specifically, {\blue the Landsat-8/9 satellites provide} broad spatial and high temporal coverage for global multispectral monitoring \cite{Landsatdata}.
More importantly, the Landsat-8/9  multispectral images (MSIs) cover a wavelength range closely aligned with that of widely used AVIRIS HSIs (i.e., about 400 nm to 2500 nm) \cite{AVIRISdata}.
As shown in Table \ref{tab:LANDSAT89}, except for the cirrus band, panchromatic band, and the two very low-resolution 100-m thermal infrared sensor (TIRS) bands (i.e., B10 and B11), the Landsat-8/9 satellite provides seven 30-m multispectral bands \cite{Landsatdata} across a wavelength range from 440 nm to 2200 nm. 
In other words, 7-band Landsat-8/9 MSIs can be regarded as a spatially and spectrally compressed counterpart of AVIRIS HSIs with 15-m spatial resolution (SpaR).
In light of the well-established compressed sensing \cite{hsu2024real,kang2024effective} and super-resolution techniques \cite{yao2026dtwstsr} in the RS community, converting medium-resolution Landsat-8/9 MSIs into high-resolution AVIRIS-level HSIs is challenging yet theoretically feasible.

This Landsat-driven global hyperspectral monitoring involves both spatial super-resolution (SpaSR, 30-m to 15-m) and the highly ill-posed spectral super-resolution (SpeSR, 7-band to 172-band).
As the duality between SpaSR and SpeSR has been mathematically derived very recently \cite[Theorem 1]{COS2A}, we refer to this challenging task as dual super-resolution (DualSR).
In the SpeSR stage, a customized interpretable (and therefore lightweight) architecture is essential.
More specifically, current SpeSR methods are primarily designed for RGB-to-31-band HSI reconstruction within the visible wavelength range \cite{hang2021spectral,RepCPSI,mei2023lightweight}.  
A representative example is the multi-stage spectral-wise transformer (MST++) \cite{MST++}, which placed first in the well-known New Trends in Image Restoration and Enhancement (NTIRE) Challenge.
However, the U-shaped spectral-wise self-attention architecture within MST++ becomes computationally prohibitive for our target 7-band to 172-band SpeSR (cf. Figure \ref{fig:Landsat_FLOPs}). 
This example indicates that, when the architecture itself lacks interpretability, the performance improvements tend to rely on deeper, more complex mechanisms, such as dense residual connections \cite{HSCNN} and multistage spatial-spectral learning \cite{MSFN}.
However, the number of model parameters in these conventional approaches \cite{MST,HRNet,HSCNN,SPECAT,MSFN,MST++} accordingly increases by multiple orders in the target AVIRIS-level SpeSR (cf. Figure \ref{fig:Landsat_FLOPs}).
Such complex architectures hinder network optimization and may lead to suboptimal solutions, as shown in Figure \ref{fig:Landsat_FLOPs}.

On the other hand, the advanced quantum SpeSR technique \cite{PRIME} has been developed to address the challenging underdetermined source separation problem \cite{MU2026TIP}.
The unitary-computing quantum features \cite{IWD_QUEEN} have demonstrated great potential for enhancing the performance of classical networks in diverse RS and EO applications, such as change detection \cite{QUEENG}, mangrove mapping \cite{QEDNET}, HSI restoration \cite{HyperQUEEN}, and denoising \cite{HyperKING}.
Nevertheless, due to the limited number of available quantum bits (qubits) and the presence of noise effects in current quantum hardware, such techniques remain difficult to deploy on this large-scale optimization framework.
The abovementioned limitations render existing SpeSR algorithms, whether classical or quantum-based, less suitable for the highly ill-posed 7-band to 172-band SpeSR.

To address this challenging task, we develop an interpretable alternating direction method of multipliers network (ADMM-Net) by incorporating the Woodbury Lemma (W-Lemma) and a spectral continuity prior.
In detail, we first formulate a convex optimization problem for the SpeSR task, where a customized spectral downsampling operator is introduced to characterize an auxiliary 86-band HSI (instead of the target 172-band HSI) and the input Landsat MSI in the data-fitting term. 
Since natural HSIs often exhibit strong spectral continuity \cite{PRIME,10684390}, the desired 172-band HSI can be obtained by spectrally upsampling the resulting auxiliary HSI.
This spectral continuity prior not only mitigates potential optimization instability (caused by a larger number of variables) and prevents mathematically sophisticated regularizers (e.g., total variation), but also improves computational efficiency. 
The experimental results presented in Section \ref{sec: experiment} will support this design.
Furthermore, when deriving the closed-form solutions for the ADMM iteration, the W-Lemma reduces the dimension of the induced matrix inversion, thereby further enhancing the closed-form implementation.
With these explicitly derived updates (without gradient-based or numerical approximations), the convex ADMM problem is unfolded into a highly interpretable ADMM-Net to solve the 7-to-172 SpeSR task.

In the SpaSR stage (30-m to 15-m), we employ an interpretable proximal gradient descent network (PGD-Net) \cite{nitanda2014stochastic} to reconstruct reliable spatial details.
Unlike conventional SpaSR approaches that primarily aim to generate visually pleasing details \cite{wu2024unsupervised, lei2026nonlocal, karwowska2023mcwesrgan,sharifi2025enhancing}, global hyperspectral monitoring requires physically faithful spatial characteristics for real-world RS applications.
Fortunately, Landsat-8/9 imagery provides high-resolution panchromatic images (cf. Table \ref{tab:LANDSAT89}), motivating us to employ a panchromatic sharpening (pansharpening) strategy to address the SpaSR problem. 
This natural advantage also explains why we employ Landsat MSI for such critical RS tasks.
Nevertheless, the formulated convex pansharpening problem remains computationally prohibitive when solved via the widely used ADMM, as the corresponding closed-form solution involves large matrix inversions (LMI) whose dimensions scale with the number of pixels (even after applying {\blue the W-Lemma}).
To resolve the dilemma, we leverage proximal gradient descent (PGD) \cite{nitanda2014stochastic} to mathematically bypass such computationally expensive LMI.
With these LMI-free implementations, we further unfold the resulting PGD-based algorithm into a lightweight PGD-Net to enable real-time computing.
The overall Landsat-driven global hyperspectral monitoring framework is referred to as PGD-ADMM interpretable network (PAINT).

Owing to its theoretically oriented and highly interpretable design, the proposed PAINT achieves a performance breakthrough of 1.94 dB (in PSNR) across diverse landscapes, including coastline/lake, mountain, farm, and city.
Furthermore, as presented in Figure \ref{fig:Landsat_FLOPs}, both computational complexity (reported in FLOPs) and memory cost (measured in network parameters) of the proposed PAINT are significantly reduced.
These strengths greatly facilitate downstream HSI-monitoring image analysis.
To demonstrate its practical applicability, we apply the PAINT-reconstructed HSI to the representative RS tasks, namely blind source separation (e.g., hyperspectral or multispectral unmixing) \cite{zhang2026ssst,MU2026TIP}.
The recovered abundance maps (i.e., spatial distributions of pure materials) are strongly consistent with the spatial patterns observed in the high-resolution Google Earth imagery. 
In addition, pixel-level hyperspectral material identification/classification has attracted considerable attention in the recent RS field \cite{chen2025dual,zhang2026central,feng2025dsnet}.
We also conduct a classification evaluation on the PAINT-reconstructed HSI.
The experimental results show that the proposed PAINT significantly advances the Landsat classification
performance from 78.98\% accuracy and 76.06\% kappa to 92.16\% accuracy and 90.98\% kappa.
These results not only demonstrate the reliability of PAINT-reconstructed spectral signatures but also highlight their practical utility in RS and EO applications.

To facilitate understanding of the proposed PAINT and its novelties, the main contributions are summarized as follows:
\begin{enumerate}
    \item In the RS and EO field, numerous critical applications rely on high SpeR and SpaR HSIs, such as AVIRIS hyperspectral imagery.
    However, direct acquisition of global HSIs remains practically infeasible due to current hardware limitations.
    A more economical alternative is to algorithmically recover them from globally available multispectral imagery (e.g., NASA's Landsat).
    This goal involves highly ill-posed 7-band to 172-band SpeSR and 30-m to 15-m SpaSR. 
    In this study, we address this challenging DualSR task using a highly interpretable (hence efficient and reliable) framework.
    To the best of our knowledge, this study is the first interpretable Landsat-driven global hyperspectral monitoring framework.
    \item To address the highly ill-posed SpeSR task, we customize an interpretable ADMM-Net that incorporates the W-Lemma and spectral continuity prior.
    In detail, we first formulate a convex SpeSR problem to derive an auxiliary 86-band HSI, rather than directly estimating the target 172-band AVIRIS-level HSI (the latter is infeasible and causes heavy network parameters).
    As natural HSIs often exhibit strong spectral continuity, the desired HSI can {\blue consequently be} obtained by applying scene-adaptive spectral shuffling on the resulting auxiliary 86-band HSI.
    This design not only bypasses complex spectral regularizers but also improves computational efficiency. 
    In addition, the W-Lemma is used to implement the induced matrix inversion in the ADMM closed-form solution.
    Hence, the convex SpeSR problem can be unfolded into a highly efficient ADMM-Net to solve the 7-to-172 SpeSR task. 
    \item In the SpaSR stage (30-m to 15-m), the LMI-free PGD-Net is further advanced for recovering reliable spatial details.
    In contrast to conventional SpaSR approaches (which focus on visually pleasing details), we employ a pansharpening strategy to address the SpaSR problem; true details are essential for EO and RS applications.
    The formulated convex pansharpening problem is solved using PDG rather than the widely used ADMM, whose closed-form solution involves a computationally expensive LMI (even after applying the W-Lemma).
    With these LMI-free updates, the resulting PGD-based algorithm is unfolded into an efficient PGD-Net.
    We refer to the overall DualSR framework as PGD-ADMM interpretable network (PAINT).
    \item Owing to its theoretically oriented and highly interpretable
    design, the proposed PAINT achieves a performance breakthrough of 1.94 dB (in PSNR) across diverse landscapes,
    including coastline/lake, mountain, farm, and city.
    Beyond quantitative evaluations, we also assess the practical applicability and identifiability \cite{EMI} of the PAINT-reconstructed HSIs using blind source separation and pixel-level classification tasks.
    Remarkably, PAINT does significantly advance the Landsat classification performance (78.98\% accuracy and 76.06\% kappa) to a new level (92.16\% accuracy and 90.98\% kappa).
    These results verify the reliability of the PAINT-reconstructed spectral signatures and demonstrate their applicability for downstream RS and EO applications.
\end{enumerate}

\bigskip

The remainder of this paper is organized as follows. 
In Section \ref{sec:thm}, we provide the detailed design philosophy of the proposed PAINT, encompassing the problem description, convex solvers, and deep network architectures.
In Section \ref{sec: experiment}, we present and discuss comprehensive experimental results, including quantitative and qualitative analyses, exhaustive case studies across diverse landscapes, and an ablation study.
Finally, Section \ref{sec:conclusion} summarizes the conclusions and future work of this study.

For better readability, the standard notations used throughout this paper are summarized below.
$\mathbb{R}^{m \times n}$, $\|\cdot\|_{F}$, and $\bm{I}$ denote the set of all the $m \times n$ real-valued matrices, the Frobenius norm, and the identity matrix, respectively. 
$\bm{A}^{T}$ and $\bm{A}^{-1}$ denote the transpose and inverse matrix of the matrix $\bm{A}$, respectively.
In this work, $\bm{Y}_{H}\in\mathbb{R}^{M\times L}$, $\bm{Y}_{L}\in\mathbb{R}^{M_{L}\times(L/r^{2})}$, $\widetilde{\bY}_L\in\mathbb{R}^{M_{L}\times L}$, and $\bm{P}\in\mathbb{R}^{M_{P}\times L}$ denote the desired HR AVIRIS-level HSI, LR Landsat-8/9 MSI, HR Landsat-8/9 MSI, and HR panchromatic image, where $M$, $L$, $M_{L}$, $M_{P}$, and $r$ represent the number of hyperspectral bands, the number of HR pixels, the number of multispectral bands, the number of panchromatic bands, and the spatial upsampling factor, respectively.
Moreover, $\bm{B}\in\mathbb{R}^{L\times(L/r^{2})}$, $\bm{D}\in\mathbb{R}^{M_{P}\times M_{L}}$, and $\widetilde{\bm{D}}\in\mathbb{R}^{M_{L}\times(M/2)}$ denote the spatial blurring/downsampling operator, the spectral downsampling operator associated with the PGD-Net, and the spectral downsampling operator associated with the ADMM-Net, respectively.

\section{The Proposed PAINT DualSR Algorithm}\label{sec:thm}

\subsection{Problem Description and Solution Outline}\label{sec:probdesc}

We target at NASA's high-standard 224-band AVIRIS HSI, covering wavelengths from about 400 nm to 2500 nm \cite{AVIRISdata}.
After removing some low-quality bands (i.e., bands 1–10, 104–116, 152–170, and 215–224) \cite{HyperKING}, we obtain the 172-band high-quality AVIRIS data $\bY_H\in\mathbb{R}^{M\times L}$, where $M:=172$ is number of HSI bands and $L$ is the number of high-resolution (HR) pixels.
However, direct acquisition of AVIRIS data is more expensive and geologically restricted to the North American \cite{AVIRISdata}.
By contrast, NASA’s longest-running Landsat satellite program provides global-wide spatiotemporal coverages, and has been broadly employed in global monitoring tasks.
We are thus motivated to computationally convert the medium-resolution Landsat-8/9 data into HR AVIRIS hyperspectral data, thereby economically obtaining datasets with comprehensive spatiotemporal coverage and superior material identifiability \cite{EMI,EMIicassp}.
However, this conversion is a highly ill-posed DualSR problem that involves both SpaSR (30-m to 15-m) and SpeSR (7-band MSI to 172-band HSI).
Therefore, compared to the prevailing CAVE-level SpeSR problem (involving only 31 visible bands) \cite{hang2021spectral,ma2022multi,qu2023unmixing}, the target inverse imaging problem (Landsat-to-AVIRIS) presents not only a highly ill-posed nature but also more significant technical challenges.
\begin{figure}[t]
\centering
\includegraphics[width=1\linewidth]{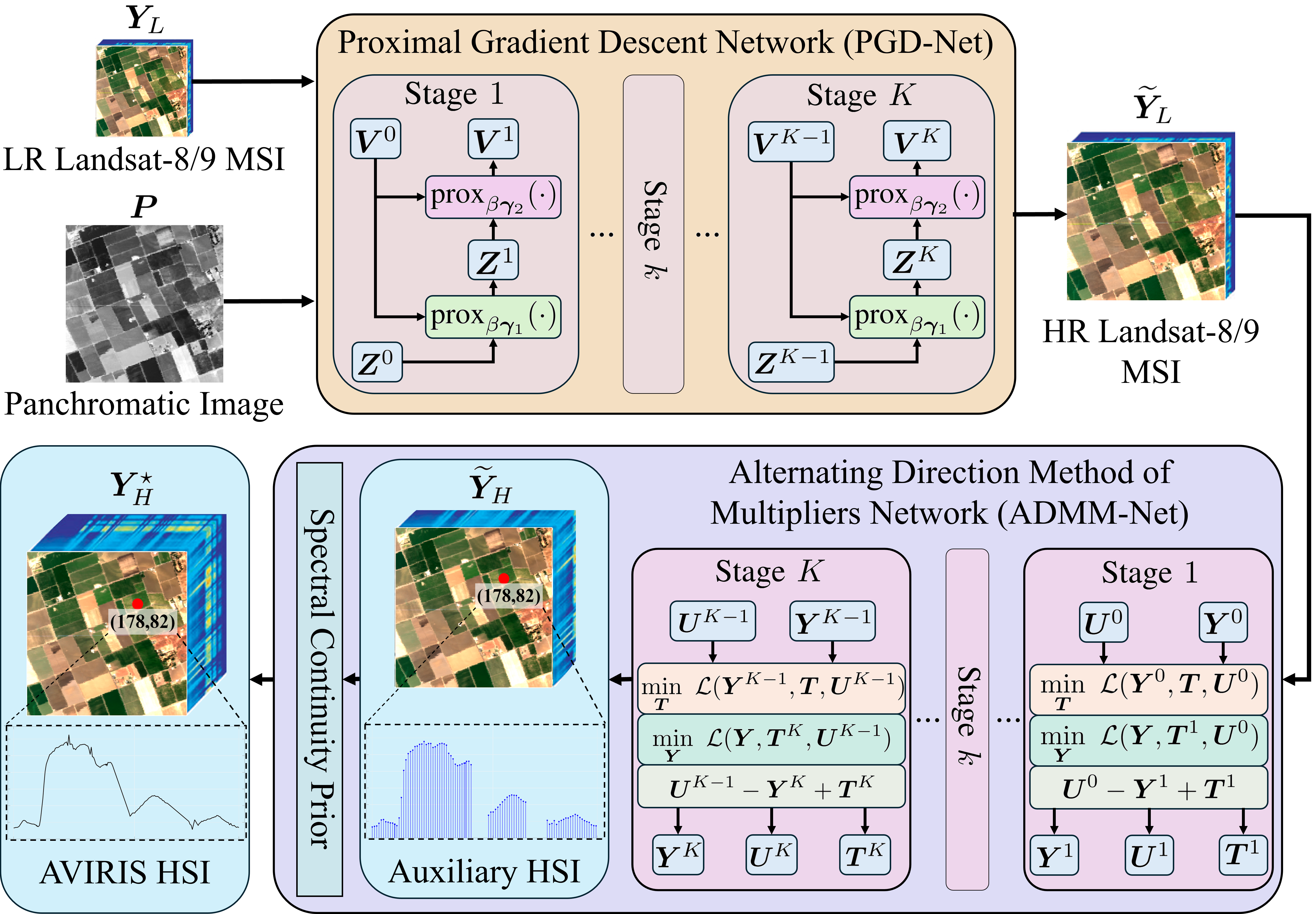}
\caption{The proposed PAINT DualSR algorithm interpretably upgrades the 7-band 30-m Landsat-8/9 MSI $\bY_L$ (i.e., B1-B7 in Table \ref{tab:LANDSAT89}) to 172-band 15-m AVIRIS-level HSI $\bY_H^\star$.
As learning spatial textures before addressing the spectral upsampling has been reported to be more effective \cite{HyperQUEEN}, PAINT first performs panchromatic sharpening, which fuses the 15-m high-resolution (HR) panchromatic information $\bP$ (i.e., B8 in Table \ref{tab:LANDSAT89}) into the 30-m low-resolution (LR) MSI $\bY_L$ for obtaining the 15-m HR MSI $\widetilde{\bY}_L$.
This fusion problem is formulated and solved by the PGD convex algorithm, whose closed-form solutions are derived and implemented as interpretable deep networks.
Then, inspired by the spectral continuity nature of typical HSI \cite{PRIME,10684390}, PAINT first learns the auxiliary 86-band HSI $\widetilde{\bY}_H$ (composed of only the odd bands of $\bY_H^\star$), from which PAINT further learns the information of even bands, followed by enforcing the spectral continuity prior on $\widetilde{\bY}_H$ to obtain the high-fidelity final solution $\bY_H^\star$.
The SpeSR from $\widetilde{\bY}_L$ to $\widetilde{\bY}_H$ is formulated as an ADMM convex problem, whose closed-form solutions are derived and used to deploy the interpretable deep architecture.
}\label{fig:overview}
\vspace{-0.3cm}
\end{figure}

Mathematically, we aim to interpretably compute the 172-band 15-m HR AVIRIS HSI $\bY_H\in\mathbb{R}^{M\times L}$ from the 7-band 30-m low-resolution (LR) Landsat-8/9 MSI $\bY_L\in\mathbb{R}^{M_L\times \left(L/r^2\right)}$ (i.e., B1-B7 in Table \ref{tab:LANDSAT89}), where $M_L:=7$ is the number of MSI bands and $r:=2$ is the spatial upsampling factor.
In other words, we consider both the SpaSR and SpeSR problems.
As learning spatial details before performing the spectral upsampling has been reported to be a more effective strategy \cite{HyperQUEEN}, the proposed PAINT algorithm first conducts SpaSR, followed by SpeSR, as outlined in Figure \ref{fig:overview}.

In the SpaSR stage, we utilize the HR information retained in the 15-m HR panchromatic band $\bP\in\mathbb{R}^{M_P\times L}$ with $M_P:=1$ (i.e., B8; cf. Table \ref{tab:LANDSAT89}), rather than simply generating visually pleasing spatial textures.
Specifically, we fuse the 15-m HR textures of $\bP$ into the 30-m LR MSI $\bY_L$, thereby obtaining the 15-m HR MSI $\widetilde{\bY}_L$ with reliable spatial details (cf. Figure \ref{fig:overview}).
To this end, we first formulate this fusion problem as a convex optimization problem.
%
In addition, as mentioned in Section \ref{sec: introduction}, we solve it using the PGD convex algorithm \cite{nitanda2014stochastic} to avoid implementing LMIs (i.e., Algorithm \ref{alg:PGD}).
We derive the PGD closed-form solutions and accordingly deploy an interpretable deep unfolding network (i.e., PGD-Net) to implement the convex PGD pansharpening algorithm.
This SpaSR stage will be detailed in Section \ref{sec:spatial}.

In the SpeSR stage, we aim to convert the 7-band HR MSI $\widetilde{\bY}_L$ into the 172-band HR HSI ${\bY}_H^\star$ (cf. Figure \ref{fig:overview}), which is a highly ill-posed inverse problem.
Inspired by the spectral continuity nature of HSI, we propose a strategy to first learn an intermediate 86-band solution $\widetilde{\bY}_H\in\mathbb{R}^{(M/2)\times L}$ (composed of only the odd bands of $\bY_H^\star$), thereby decoupling the challenging SpeSR into two sub-stages, one transforming $\widetilde{\bY}_L$ to $\widetilde{\bY}_H$ (cf. Figure \ref{fig:overview}) and the other converting $\widetilde{\bY}_H$ to ${\bY}_H^\star$ (cf. Figure \ref{fig:overview}).
This also echoes the convex/deep (CODE) small-data learning theory \cite{CODE}, which advocates first learning an intermediate rough solution via deep learning, followed by refining it to obtain the final solution via convex optimization.
This strategy has received considerable success in various challenging hyperspectral inverse problems \cite{CODEHCD,CODEIF}.
In the first sub-stage (from $\widetilde{\bY}_L$ to $\widetilde{\bY}_H$), the 7-to-86 SpeSR is formulated as an ADMM problem (cf. Algorithm \ref{alg:ADMM}), whose closed-form solutions are derived with {\blue the W-Lemma} to implement {\blue interpretable} deep network elements.
In the second sub-stage (from $\widetilde{\bY}_H$ to ${\bY}_H^\star$), PAINT further learns the information of even bands to compensate $\widetilde{\bY}_H$, followed by the spectral shuffling with the odd bands to obtain the target HR HSI $\bY_H^\star$ (cf. Figure \ref{fig:overview}).
More details are presented in Section \ref{sec:spectral}.

Beyond the model architecture design, which mathematically bypasses the induced LMI via convex optimization theory.
It is worth noting that acquiring high-quality AVIRIS-Landsat image pairs over identical geographical regions and time periods remains significantly challenging, thereby limiting the availability of training data.
This stems primarily from spatial resolution discrepancies between the sensors (i.e., non-integer multiples), the frequently encountered cloudy areas, and the inconsistencies in the flying trajectories \cite{AVIRISdata, Landsatdata}.
The high interpretability of the proposed PAINT algorithm also effectively mitigates the need for a large dataset (trained using only hundreds of pairs) to solve the highly ill-posed target DualSR problem.
The pretrained PAINT demonstrates strong generalization ability across diverse landscapes, as presented in Section \ref{sec: experiment}.
%



\subsection{LMI-Free PGD Spatial Super-resolution Network}\label{sec:spatial}

Due to the limited spatial resolution of Landsat-8/9, straightforward SpeSR inevitably yields spatially coarse HSI and suffers from {\blue a serious} mixed-pixel phenomenon \cite{HyperCSI}, thereby rendering its applicability in real-world scenarios.
However, conventional SpaSR approaches typically rely on learnable upsampling architectures \cite{reviewer1,reviewer2}, which infer spatial details solely from LR inputs without explicit HR guidance, resulting in suboptimal reconstruction performance.
Moreover, conventional pansharpening models are not explicitly designed for downstream SpeSR tasks, which can cause incompatibility with subsequent reconstruction modules.
To address this limitation, the complementary spatial and spectral information from the HR panchromatic band and the LR Landsat MSI bands are fused to reconstruct the HR Landsat imagery, and this customized fusion network will be trained with the ADMM SpeSR network (cf. Section \ref{sec:spectral}) in an end-to-end manner.

To infer the HR Landsat imagery $\widetilde{\bY}_L \in \mathbb{R}^{M_L\times L}$ from HR panchromatic band $\bP\in \mathbb{R}^{M_P\times L}$ and LR MSI bands $\bY_L\in \mathbb{R}^{M_L\times\left(L/r^2\right)}$, a natural pansharpening criterion can be concisely formulated as 
\begin{equation}\label{prob:PGD}
\widetilde{\bY}_L
:= 
\arg\min_{\bZ}
\left\| \bY_L - \bZ\bB\right\|_F^2
+ \left\| \bP - \bD\bZ\right\|_F^2
+ \beta\bm\gamma(\bZ),
\end{equation}
where $\bB\in \mathbb{R}^{L\times\left(L/r^2\right)}$ is the spatial blurring operator, $\bD\in \mathbb{R}^{M_P\times M_L}$ is the spectral downsampling operator, and $\bm\gamma(\cdot)$ represents an implicit deep regularizer with strength controlled by the regularization parameter $\beta>0$.
As the deep network itself already serves as an image prior \cite{DIP}, we only need an implicit regularization for deep unfolding, and this is proven highly effective for related super-resolution tasks \cite{COS2A}.
Moreover, the regularization parameter $\beta$ corresponds to the step size in the PGD algorithm \cite{PGD_CY}, and is hence set as the commonly used value $\beta:=0.001$ under the deep unfolding framework.
The first and second terms of \eqref{prob:PGD} encourage the spatially blurred version and the spectrally downsampled version of $\widetilde{\bY}_L\equiv\bZ$ to fit the observable LR image $\bY_L$ and panchromatic image $\bP$, respectively.

If we solve \eqref{prob:PGD} using ADMM, the SpaSR will induce a large matrix inversion with the dimension proportional to the number of pixels, i.e., $L$, which is very large.
We judiciously come up with solving \eqref{prob:PGD} using PGD algorithm as an alternative, thereby bypassing the matrix inversion, as will be demonstrated below.
To this end, we introduce an auxiliary variable $\bV\in \mathbb{R}^{M_L\times L}$, and reformulate the criterion \eqref{prob:PGD} as 
\begin{equation} \label{prob:PGD_reform}
\min_{\bZ,\bV}
\left\| \bY_L - \bZ\bB\right\|_F^2
+ \left\| \bP - \bD\bV\right\|_F^2
+ \frac{\rho}{2}\left\| \bZ - \bV\right\|_F^2
+ \beta\bm\gamma(\bZ),
\end{equation}
%
where $\rho>0$ is the penalty parameter.
As an advantage of deep unfolding, $\rho$ can be set as a learnable parameter.
%
%
%
%
Then, the PGD algorithm \cite{parikh2014proximal,nitanda2014stochastic} solves \eqref{prob:PGD_reform} by the iterative updates,
\begin{align}
\bZ^{k+1}
&=\textrm{prox}_{\beta\bm\gamma_1}(\bZ^k+\beta\nabla f_1(\bZ^k\mid\bV^k)), \label{eq:PGD-Z}
\\
\bV^{k+1}
&=\textrm{prox}_{\beta\bm\gamma_2}(\bV^k+\beta\nabla f_2(\bV^k\mid \bZ^{k+1})) \notag 
\\
&=\bV^k+\beta\nabla f_2(\bV^k\mid \bZ^{k+1}), \label{eq:PGD-V}
\end{align}
where $\textrm{prox}_{\gamma}(\bv)\triangleq\arg\min_{\bx} \gamma(\bx)+\frac{1}{2}\|\bx-\bv\|_2^2$ denotes the proximal operator associated with the function $\gamma$ \cite{parikh2014proximal}, as detailed in Algorithm \ref{alg:PGD}.
Here, $\bm\gamma_1(\bZ)\triangleq\bm\gamma(\bZ)$ and $f_1(\bZ\mid\bV)\triangleq\left\| \bY_L - \bZ\bB\right\|_F^2
+ \frac{\rho}{2}\left\| \bZ - \bV\right\|_F^2$ are the $\bZ$-dependent non-differentiable and differentiable terms in \eqref{prob:PGD_reform}, respectively.
Similarly, $\bm\gamma_2(\bv)\triangleq 0$ and $f_2(\bV\mid\bZ)\triangleq \left\| \bP - \bD\bV\right\|_F^2+ \frac{\rho}{2}\left\| \bZ - \bV\right\|_F^2$ are the $\bV$-dependent non-differentiable and differentiable terms in \eqref{prob:PGD_reform}, respectively.
Note that the proximal operator in \eqref{eq:PGD-V} can be simplified as the identity matrix (i.e., $\textrm{prox}_{\beta\bm\gamma_2}(\bv)=\textrm{prox}_{0}(\bv)
=\arg\min_{\bx} \frac{1}{2}\|\bx-\bv\|_2^2
=\bv$).

\begin{figure*}[t]
\centering
\includegraphics[width=1\linewidth]{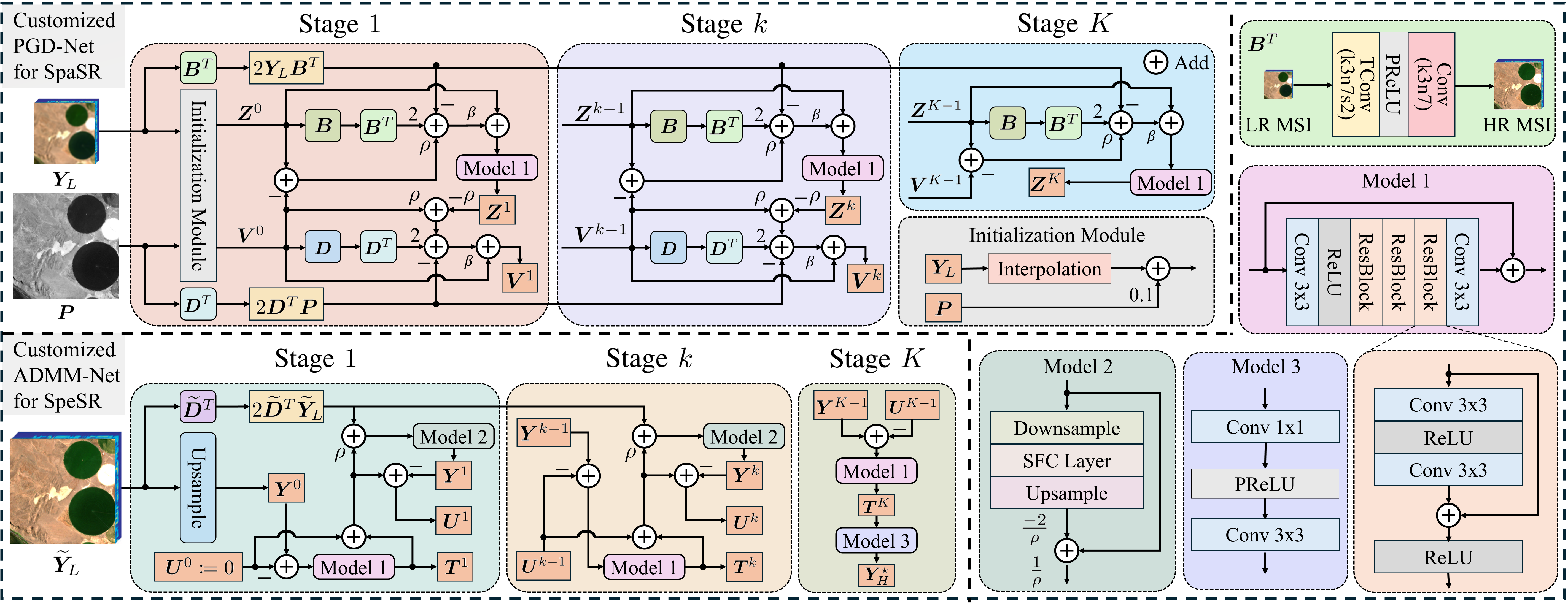}
%
\caption{The proposed PAINT DualSR algorithm provides an interpretable transform from the 7-band 30-m Landsat-8/9 MSI $\bY_L$ to the 172-band 15-m AVIRIS-level HSI $\bY_H^{\star}$.
In the PGD SpaSR network, PAINT first derives the HR MSI $\widetilde{\bY}_L$ from the LR MSI $\bY_L$ and the HR panchromatic image $\bP$ using the deep unfolding network of PGD algorithm (i.e., Algorithm \ref{alg:PGD}). 
Specifically, the initialization module is applied to initialize the variables $\bZ$ and $\bV$ by roughly injecting the HR panchromatic information of $\bP$ into the LR MSI $\bY_L$.
Then, $\bB$ (spatial blurring) and $\bB^T$ (spatial upsampling) operators are implemented by the 2-by-2 average pooling and transposed convolution (TConv) network, respectively.
In the TConv network of $\bB^T$, the network component ``k($b$)n($c$)s($\ell$)'' denotes the $c$-channel convolution layer with stride $\ell$ and $b\times b$ kernels; the network component ``k($b$)n($c$)'' is defined in a similar way.
Additionally, $\bD$ and $\bD^{T}$ are trainable linear transformations, representing the spectral downsampling and upsampling operations, respectively.
Finally, as residual-in-residual architecture is proven to be a strong denoiser, Model 1 adopts the residual-in-residual network to implement the proximal operator (mathematically equivalent to a denoiser \cite{CVXbookCLL2016,COS2A}) induced by the implicit regularizer for the ill-posed pansharpening problem, thereby returning the HR Landsat image $\widetilde{\bY}_L:=\bZ^K$, as detailed in Algorithm \ref{alg:PGD}.
Next, in the ADMM SpeSR network, we adopt the ADMM algorithm (i.e., Algorithm \ref{alg:ADMM}), inducing three basic models (i.e., Model 1, Model 2, and Model 3) to unfold the deep network.
Model 1 again implements the proximal denoiser using the residual-in-residual deep architecture.
Model 2 employs the symmetricity of ${\bm\Phi}$ [as well as the physical meanings of $\widetilde{\bD}$ (spectral downsampling) and $\widetilde{\bD}^T$ (spectral upsampling)] to implement the operator ``$\widetilde{\bD}^T{\bm\Phi}\widetilde{\bD}$'', and uses a shortcut connection to implement the identity operator ``$\bI$''.
Model 3 is designed to promote the spectral continuity nature of typical HSIs, thereby converting the auxiliary $\widetilde{\bY}_H$ into the target HSI $\bY_H^\star$.} 
\label{fig:network}
\end{figure*}

Accordingly, the closed-form updating rules under the PGD algorithm are derived as follows, i.e.,
\begin{align}
\bZ^{k+1}
&= 
{\bm\Pi}
[\bZ^k+\beta(
2(\bZ^k\bB\bB^T-\bY_L\bB^T)
+
\rho(\bZ^k-\bV^k)
)], \label{eq:PGD-Z-sol}
\\
\bV^{k+1}
&=
\bV^k+\beta(2\bD^T\bD\bV^k-2\bD^T\bP+
\rho(\bV^k-\bZ^{k+1})), \label{eq:PGD-V-sol}
\end{align}
where we define ${\bm\Pi}(\cdot)\triangleq \textrm{prox}_{\beta\bm\gamma_1}(\cdot)=\textrm{prox}_{\beta\bm\gamma}(\cdot)$ as the proximal operator in \eqref{eq:PGD-Z} for notational simplicity.
The corresponding PGD SpaSR networks of \eqref{eq:PGD-Z-sol}-\eqref{eq:PGD-V-sol} are designed in the upper part of Figure \ref{fig:network}, where the model-order is empirically set as $K:=4$ \cite{COS2A}.

\begin{algorithm}[t]
\caption{PGD Spatial Super-resolution Algorithm}
\begin{algorithmic}[1]\label{alg:PGD}
\STATE {\bf Given}
LR Landsat-8/9 MSI $\bY_L$, and HR panchromatic image $\bP$. 

\STATE 
Initialize $k:=0$ and $(\bZ^0,\bV^0)$ using the initialization module (cf. Figure \ref{fig:network}).

\REPEAT

\STATE 	
Update 
$\bZ^{k+1}
=
\text{prox}_{\beta\bm\gamma_1}(\bZ^k+\beta\nabla f_1(\bZ^k\mid\bV^k))$,

\STATE 		
Update 
$\bV^{k+1}
=
\text{prox}_{\beta\bm\gamma_2}(\bV^k+\beta\nabla f_2(\bV^k\mid\bZ^{k+1}))$,

\STATE 
$k:=k+1$,

\UNTIL the predefined stopping criterion is met.

\STATE {\bf Output} the HR MSI $\widetilde{\bY}_L:=\bZ^k$.
\end{algorithmic}
\end{algorithm}

In each stage $k$ of the PGD architecture, we first form the partial image ``$\mathcal{IM}_1\triangleq 2(\bZ^k\bB\bB^T-\bY_L\bB^T)+\rho(\bZ^k-\bV^k)$'' through the network deployment in Figure \ref{fig:network}, followed by computing the full image ``$\mathcal{IM}_2\triangleq \bZ^k+\beta~\!\mathcal{IM}_1$'' (cf. Figure \ref{fig:network}), which is then fed into the proximal operator ${\bm\Pi}(\cdot)$ (cf. \eqref{eq:PGD-Z-sol}) that is implemented by Model 1 (cf. Figure \ref{fig:network}).
Since the physical meaning of the proximal operator is nothing but a regularized denoiser \cite{CVXbookCLL2016,parikh2014proximal}, $\bZ^{k+1}$ can be considered as a denoised version of $\mathcal{IM}_2$ according to \eqref{eq:PGD-Z-sol}.
In other words, we recover the clean image $\bZ^{k+1}$ with some desired property specified by the deep denoising regularizer $\bm\gamma$ in $\bm\Pi$, for which we adopt the simple residual-in-residual architecture that is proven highly effective for denoising \cite{COS2A,DIP,zhang2021plug}, as detailed in Model 1 of Figure \ref{fig:network}.
The blurring operator $\bB$ is implemented by the average pooling; $\bB^T$ is regarded as the spatial upsampling operator, for which the learnable transpose convolution (TConv) is known to be effective \cite{CODE}, as detailed in Figure \ref{fig:network} for reproducibility.
Following a similar strategy, $\bV^{k+1}$ can be deeply unfolded via \eqref{eq:PGD-V-sol}, where $\bD$ and $\bD^T$ are also set as learnable parameters, as deployed in Figure \ref{fig:network}.
In the PGD architecture, the first stage (i.e., Stage 1) has an additional initialization module, which roughly injects the spatial details of $\bP$ into $\bY_L$ as illustrated in Figure \ref{fig:network}.
As for the final stage (i.e., Stage $K$) of the PGD network, it is greatly simplified as it only needs to return $\bZ^K\equiv \widetilde{\bY}_L$ (cf. Algorithm \ref{alg:PGD}), as detailed in Figure \ref{fig:network}.
Therefore, we have completed the deep unfolding network design of the PGD convex algorithm (i.e., Algorithm \ref{alg:PGD}).

\subsection{{Efficient ADMM Spectral Super-resolution Network}}\label{sec:spectral}

After obtaining the HR MSI $\widetilde{\bY}_L\in\mathbb{R}^{M_L\times L}$, the next step is to spectrally super-resolve it into the HR HSI $\bY^{\star}_H\in\mathbb{R}^{M\times L}$ (cf. Figure \ref{fig:overview}).
To mitigate the computational burden of the unfolded architecture, we first reconstruct an auxiliary rough solution $\widetilde{\bY}_H\in\mathbb{R}^{(M/2)\times L}$ (composed of only the odd bands of $\bY^{\star}_H$).
Subsequently, we exploit the spectral continuity prior, implemented by a lightweight convolutional neural network (CNN), to obtain the target solution $\bY^{\star}_H$ (cf. Figure \ref{fig:overview}).
This strategy also echoes the CODE learning theory (learning a deep rough solution first) \cite{CODE}, as discussed in Section \ref{sec:probdesc}, and is ensured to work effectively owing to the spectral continuity of typical hyperspectral pixels.
Furthermore, as Landsat-to-AVIRIS conversion is a highly ill-posed inverse problem, we will apply Woodbury matrix theory to further reduce the burden of the deep learning task, as detailed later.

A natural criterion of estimating $\widetilde{\bY}_H$ from $\widetilde{\bY}_L$ can be formulated as the following optimization problem, i.e.,
\begin{equation}\label{prob:DIP}
\widetilde{\bY}_H
:= 
\arg\min_{\bY}
\left\| \widetilde{\bY}_L -  \widetilde{\bD} \bY \right\|_F^2
+ \mathrm{DIP}(\bY), 
\end{equation}
where $\widetilde{\bD}\in\mathbb{R}^{M_L\times (M/2)}$ is the spectral downsampling operator, and $\textrm{DIP}(\cdot)$ is again an implicit deep image prior (DIP) \cite{DIP,PRIME,MU2026TIP}.
This problem can be equivalently reformulated as $\min_{\bY=\bT}\|\widetilde{\bY}_L-\widetilde{\bD}\bY\|_F^2+\textrm{DIP}(\bT)$, and solved using the ADMM primal-dual updating rules \cite{CVXbookCLL2016}, i.e.,
\begin{align}
\bT^{k+1}
&\in\arg\min_{\bT}~\mathcal{L}(\bY^{k},\bT,\bU^k),
\label{eq:ADMM-T}
\\
\bY^{k+1}
&\in\arg\min_{\bY}~\mathcal{L}(\bY,\bT^{k+1},\bU^k),
\label{eq:ADMM-Y}
\\
\bU^{k+1}
&:=\bU^k-\bY^{k+1}+\bT^{k+1},
\label{eq:ADMM-U}
\end{align}
where \eqref{eq:ADMM-T}-\eqref{eq:ADMM-Y} are primal updates, \eqref{eq:ADMM-U} is called dual update for the dual variable $\bU$ in the ADMM theory, and 
$\mathcal{L}(\bY,\bT,\bU)
\triangleq 
\|\widetilde{\bY}_L-\widetilde{\bD}\bY\|_F^2
+\textrm{DIP}(\bT)
+\frac{\rho}{2} \|\bY-\bT-\bU\|_F^2$ is the augmented Lagrangian of \eqref{prob:DIP} \cite{CVXbookCLL2016}.
This is detailed in Algorithm \ref{alg:ADMM}, where the penalty parameter $\rho>0$ is set as a learnable parameter in the ADMM unfolding network, and $\bm 0$ denotes the zero matrix for initializing the dual variable $\bU^0$.

According to the definition of proximal operator (cf. Section \ref{sec:spatial}) and the definition of $\mathcal{L}(\bY,\bT,\bU)$, the closed-form solutions of $\bT^{k+1}$ and $\bY^{k+1}$ can be derived as 
\begin{align}
\bT^{k+1}
&=
\textrm{prox}_{\frac{1}{\rho}\textrm{DIP}}({\bY}^k-\bU^k), \label{prob:T-Update}
\\
\bY^{k+1}
&=
(2\widetilde{\bD}^T\widetilde{\bD}+\rho \bI)^{-1}\left(2\widetilde{\bD} ^T \widetilde{\bY}_L+\rho(\bT^{k+1}+\bU^k)\right). \label{prob:Y-Update}
\end{align}
The ADMM-Net (cf. the lower part of Figure \ref{fig:network}) prepares the two input images, denoted as ``$\mathcal{IM}_3\triangleq {\bY}^k-\bU^k$'' (cf. \eqref{prob:T-Update}) and ``$\mathcal{IM}_4\triangleq 2\widetilde{\bD}^T \widetilde{\bY}_L+\rho(\bT^{k+1}+\bU^k)$'' (cf. \eqref{prob:Y-Update}).
The model-order is empirically set to $K:=3$.
Then, ``$\mathcal{IM}_3$'' is fed into Model 1 to implement the proximal operator in \eqref{prob:T-Update}.
Recall that the physical meaning of proximal mapping is image denoising, and Model 1 adopts the residual-in-residual architecture to implement the denoiser, as detailed in Section \ref{sec:spatial}.
Therefore, $\bT^{k+1}$ (i.e., the output of Model 1) can be considered as the denoised version of ``$\mathcal{IM}_3$'', as graphically illustrated in Figure \ref{fig:network}.

Similarly, after computing $\bT^{k+1}$, the ADMM-Net can form the image ``$\mathcal{IM}_4$'' (defined above).
However, the matrix inversion of \eqref{prob:Y-Update} has a dimension depending on the number of hyperspectral bands, which is too large.
Therefore, we adopt the Woodbury matrix inversion lemma \cite{CVXbookCLL2016} to equivalently represent the matrix inversion as
\begin{equation}\label{prob:Y_H-Update}
(2\widetilde{\bD}^T\widetilde{\bD}+\rho \bI)^{-1}
= 
\frac{1}{\rho}\left(\bI - \frac{2}{\rho}\widetilde{\bD}^T
{\bm\Phi}
\widetilde{\bD}\right),
\end{equation}
where ${\bm\Phi}\triangleq (\bI+\frac{2}{\rho}\widetilde{\bD}\widetilde{\bD}^T)^{-1}$ is a small $7\times 7$ symmetric matrix, thereby successfully avoiding the dilemma of LMI.
Furthermore, we implement the operator of \eqref{prob:Y_H-Update} using Model 2, where the shortcut connection corresponds to the identity matrix $\bI$ {\blue on the right-hand} side of \eqref{prob:Y_H-Update} (cf. Figure \ref{fig:network}).
Also, in Model 2, the downsampling layer, upsampling layer, and symmetric fully connected (SFC) layer corresponds to $\bD$, $\bD^T$, and the symmetric matrix ${\bm\Phi}$ in \eqref{prob:Y_H-Update}, respectively, and are all set as learnable network parameters, as detailed in Figure \ref{fig:network}.
Then, based on \eqref{prob:Y-Update} and \eqref{prob:Y_H-Update}, $\bY^{k+1}$ can be obtained by feeding ``$\mathcal{IM}_4$'' into Model 2, as deployed in Figure \ref{fig:network}.

\begin{algorithm}[t]
\caption{ADMM Spectral Super-resolution Algorithm}
\begin{algorithmic}[1]\label{alg:ADMM}
\STATE {\bf Given}
HR Landsat-8/9 image $\widetilde{\bY}_L$, and $\rho>0$. 

\STATE 
Initialize $k:=0$ and $\bU^0:=\bm 0$, and spectrally upsample $\widetilde{\bY}_L$ to initialize ${\bY}^0$ (cf. Figure \ref{fig:network}).

\REPEAT

\STATE 	
Update $\bT^{k+1}\in\arg\min_{\bT}~\mathcal{L}(\bY^{k},\bT,\bU^k)$,

\STATE 		
Update $\bY^{k+1}\in\arg\min_{\bY}~\mathcal{L}(\bY,\bT^{k+1},\bU^k)$,

\STATE 
Update $\bU^{k+1}:=\bU^k-\bY^{k+1}+\bT^{k+1}$,

\STATE 
$k:=k+1$,

\UNTIL the predefined stopping criterion is met.

\STATE {\bf Output} the HR HSI $\widetilde{\bY}_H:=\bT^k$.
\end{algorithmic}
\end{algorithm}	

Similar to the PGD network designed in Section \ref{sec:spatial}, the first stage (i.e., Stage 1) of the ADMM network has an additional initialization layer, and the final stage (i.e., Stage $K$) of the ADMM network is greatly simplified as it only needs to return $\bT^K\equiv \widetilde{\bY}_H$ (cf. Algorithm \ref{alg:ADMM}), as detailed in Figure \ref{fig:network}.
Note that in the final stage (i.e., Stage $K$), Model 3 further transforms the rough solution $\bT^K\equiv \widetilde{\bY}_H\in\mathbb{R}^{(M/2)\times L}$ (composed of only the odd bands) into the final solution $\bY_H^\star\in\mathbb{R}^{M\times L}$ using simple lightweight CNN, which is based on the spectral continuity prior of the hyperspectral bands (cf. Figure \ref{fig:network}).
Once Model 1 and Model 2 return $\bT^{k+1}$ and $\bY^{k+1}$, respectively, the ADMM-Net automatically updates of the dual variable $\bU^{k+1}$ in \eqref{eq:ADMM-U}, as can be seen from Figure \ref{fig:network}.
Therefore, we have completed the deep unfolding network design of the ADMM SepSR algorithm (i.e., Algorithm \ref{alg:ADMM}), thereby allowing us to interpretably convert historical Landsat-8/9 data into their corresponding high-standard HR AVIRIS counterpart $\bY_H^\star$ for the first time.

\begin{table*}[t]
\centering
\caption{Detailed configurations of the collected real AVIRIS HSI and Landsat-8/9 MSI pairs.
The AVIRIS HSIs were acquired from \cite{AVIRISdata}, while their corresponding Landsat-8/9 MSIs were obtained from \cite{Landsatdata}.
All data pairs were acquired over California, USA, during May 2024, June 2024, and May 2025.
More details regarding the data acquisition and preprocessing are provided in Section \ref{sec:experimental setting}.}
\label{tab:data_combined}
\scalebox{1.05}{
\begin{tabular}{|c|c|c|c|c|c|c|}
\hline
\multirow{2}{*}{Site Name} & \multicolumn{3}{c|}{AVIRIS} &  \multicolumn{3}{c|}{Landsat-8/9} \\
\cline{2-7}
& GSD (m) & Spectral Range ($\mu$m) & Image Size & GSD (m) & Spectral Range ($\mu$m) & Image Size \\
\hline
C210 & 16.0 & 0.4–2.5 & 8570$\times$877 & 30 & 0.435–2.294 & 4571$\times$468 \\
C216 & 16.2 & 0.4–2.5 & 12552$\times$1118 & 30 & 0.435–2.294 & 6778$\times$604 \\
C256 & 14.7 & 0.4–2.5 & 21581$\times$1339 & 30 & 0.435–2.294 & 10575$\times$656 \\
C279 & 15.8 & 0.4–2.5 & 18906$\times$1632 & 30 & 0.435–2.294 & 9957$\times$860 \\
SC7 & 16.0 & 0.4–2.5 & 26087$\times$1617 & 30 & 0.435–2.294 & 13913$\times$862\\
BA52 & 16.5 & 0.4–2.5 & 16488$\times$769 & 30 & 0.435–2.294 & 9069$\times$423 \\
\hline
\end{tabular}
}
\end{table*}

\section{Experimental Results and Discussion}\label{sec: experiment}
%
This section is organized below to facilitate a better understanding.
Section \ref{sec:experimental setting} first presents experimental settings, including the preparation of AVIRIS/Landsat data pairs and training configurations.
In Section \ref{sec:Results}, qualitative and quantitative evaluations are conducted to demonstrate the effectiveness of our PAINT algorithm.
Section \ref{sec: Classification} performs a case study on the classification task by employing the Landsat-level MSI, PAINT-reconstructed HSI, and real-world HSI as various input representations, and subsequently compares the classification performance.
Section \ref{sec: BSS} further extends the case study to blind source separation using real-world Landsat MSI acquired over California, USA, to demonstrate the practical utility of our PAINT algorithm.
Section \ref{sec: Discussion} provides an in-depth discussion regarding the reconstruction reliability and robustness of the simulation protocol.
Finally, Section \ref{sec:ablation} presents a comprehensive ablation study to systematically evaluate the contribution of each component in the proposed PAINT framework.

\subsection{Experimental Setting}\label{sec:experimental setting}
Since this Landsat-driven global hyperspectral monitoring task remains a pioneering research topic, a generally accepted experimental protocol and a public dataset are scarce.
To address this problem, a customized protocol is established to facilitate qualitative and quantitative assessments.
Due to space limitations, the corresponding details are presented in Supplementary Note 1.
Following the protocol, the paired AVIRIS/Landsat reference samples ($\bY_H$, $\widetilde{\bY}_H$, $\widetilde{\bY}_L$, $\bY_L$, $\bP$) can be obtained for qualitative and quantitative evaluations.

To provide guidance for future studies in the RS and EO community, we also endeavor to acquire high-quality AVIRIS/Landsat data pairs, despite the restricted spatiotemporal alignment.
Specifically, due to the different acquisition conditions and flying trajectories between AVIRIS and Landsat sensors, the data pairs exhibit resolution discrepancies and spatial misalignment, which hinder training stability.
Therefore, we carefully design the protocol of real-data preprocessing to address these issues.
Owing to limited space, the associated descriptions are summarized in Supplementary Note 2.
The case studies presented in the following sections are based on a model fine-tuned on a real dataset (collected using our protocol) to assess the practical utility of our PAINT algorithm.
Additional real-data information, including acquisition sites, ground sampling distances (GSD), spectral ranges, and spatial sizes, is summarized in Table \ref{tab:data_combined}.

The loss function for training PAINT is presented below. 
Let $(\bY_H, \bY^{\star}_H)$, $(\widetilde{\bY}_H,\widetilde{\bY}^{\star}_H)$, and $(\widetilde{\bY}_L,\widetilde{\bY}^{\star}_L)$ denote the (reference, estimation) pairs of AVIRIS HSI, auxiliary HSI, and HR Landsat-8/9 MSI, respectively, the data-driven loss function $\mathcal{L}_D$ of the proposed PAINT can be explicitly expressed as,
\begin{align*}
    \mathcal{L}_D:=&\|{\bY}_H-{\bY}^{\star}_H\|_1+\|\widetilde{\bY}_L-\widetilde{\bY}^{\star}_L\|_1
    \\
    &+\lambda_1\|\widetilde{\bY}_H-\widetilde{\bY}^{\star}_H\|_1+\lambda_2\text{SAM}(\bY_H,{\bY}^{\star}_H),
\end{align*}
where $\|\cdot\|_1$ and $\text{SAM}(\cdot)$ represent the $\ell_1$-loss function \cite{adam} and the function of spectral angle mappers (SAM) \cite{SISHY}, respectively. 
Additionally, a prior-aided loss function $\mathcal{L}_P$ is further incorporated  to promote the spectral-spatial smoothness, i.e.,
\begin{align*}
    \mathcal{L}_P:=\lambda_3\text{TV}_{\text{spe}}({\bY}^{\star}_H)+\lambda_4\text{TV}_{\text{spa}}({\bY}^{\star}_H),
\end{align*}
where $\text{TV}_{\text{spe}}(\cdot)$ and $\text{TV}_{\text{spa}}(\cdot)$ represent the spectral total variation (TV) and spatial TV \cite{zhengTVloss}, respectively.
In this study, the trade-off parameters are set to $\lambda_1:=10^{-5}$, $\lambda_2:=0.2$, $\lambda_3:=10^{-3}$, and $\lambda_4:=10^{-8}$.  
Consequently, the overall loss function of the proposed PAINT can be obtained by combining $\mathcal{L}_D$ and $\mathcal{L}_P$.
The exhaustive training settings of the proposed PAINT and benchmark methods are collectively summarized in Supplementary Note 3 for conciseness.

\subsection{Qualitative and Quantitative Analysis}\label{sec:Results}  
\begin{figure*}[t]
\centering
\includegraphics[width=1.0\linewidth]{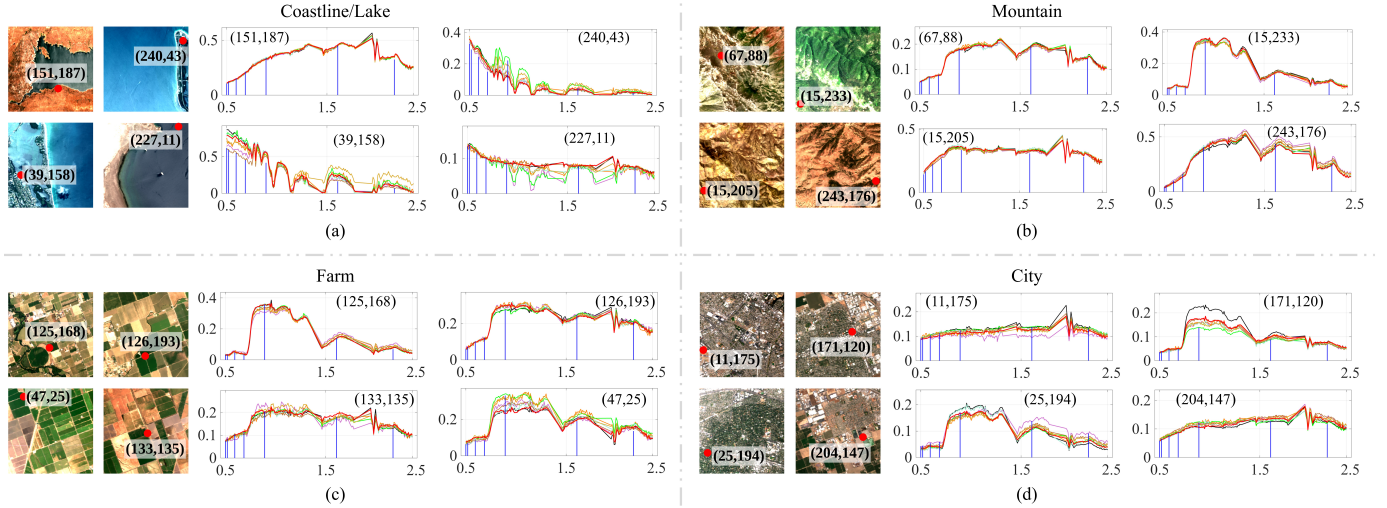}
\caption{Qualitative assessment of spectral fidelity across (a) coastline/lake, (b) mountain, (c) farm, and (d) city landscapes.
For each landscape, four pixels are uniformly selected to evaluate the spectral fidelity of the HSIs reconstructed by MST++, SPECAT, JSSUNet, SSUNet, COS2A, SepSSJSR, and the proposed PAINT, where the corresponding spatial locations of these pixels are marked by red dots in the true-color compositions of the reference AVIRIS HSIs.
The vertical and horizontal axes of the spectral curve plots denote reflectance and wavelength, respectively.
In addition, the black curve and blue discrete pulses denote the spectral signatures of the reference HSIs and their Landsat-8/9 counterparts, respectively; while the mint green, orange, brown, yellow, green, purple, and red curves represent the corresponding hyperspectral pixels reconstructed using MST++, SPECAT, JSSUNet, SSUNet, COS2A, SepSSJSR, and PAINT, respectively.
Obviously, the red curves (i.e., PAINT) best match the black curves (i.e., ground truth).
We refer readers to Supplementary Figure 1 for more in-depth qualitative comparisons.
Further analyses are summarized in Section \ref{sec:Results}.
}
\label{fig:spec_curve}
\end{figure*}

In this section, we assess the effectiveness of the proposed PAINT by comparing it with representative benchmark methods, including the MST++ \cite{MST++}, SPECAT \cite{SPECAT}, JSSUNet \cite{JSSUNet}, SSUNet \cite{SSUNet}, COS2A \cite{COS2A}, and SepSSJSR \cite{SepSSJSR}.
Specifically, MST++ and SPECAT are transformer-based frameworks that demonstrate strong performance in the prevailing RGB-to-31-band hyperspectral reconstruction tasks.
In particular, MST++ won first place in the prestigious NTIRE challenge \cite{MST++}.
On the other hand, SSUNet is a novel residual-attention-based method that integrates residual blocks and attention mechanisms to extract complementary spatial and spectral features.
As the design of PAINT also echoes the CODE small-data learning theory (as mentioned in Section \ref{sec:probdesc}), we incorporate the CODE-based framework (i.e., COS2A) for comprehensive comparisons.
Furthermore, JSSUNet is also a deep unfolding method that explicitly decomposes the reconstruction process into the SpaSR and SpeSR subtasks, making it a particularly relevant baseline for evaluating the effectiveness of PAINT.
In addition, SepSSJSR, for the first time, adopts a full CNN-based architecture for this challenging DualDR task.
To demonstrate the advantage of our interpretable design, we also employ SepSSJSR as the baseline method in the following comparisons.
To assess the robustness and generalization capability of these methods, evaluations were conducted across four distinct scenes, including coastline/lake, mountain, farm, and city, as shown in Figure \ref{fig:spec_curve}.
The following will present a comprehensive analysis and in-depth discussion.
\begin{table*}[t]
\footnotesize
\centering
\caption{Quantitative comparisons of benchmark algorithms, including MST++ \cite{MST++}, SPECAT \cite{SPECAT}, JSSUNet \cite{JSSUNet}, SSUNet \cite{SSUNet}, COS2A \cite{COS2A}, SepSSJSR \cite{SepSSJSR}, and the proposed PAINT. 
The symbol ``$\uparrow$'' (resp., ``$\downarrow$'') means that the higher (resp., lower) value is better, and accordingly those best performances are highlighted using boldfaced numbers.
The computational time is reported in \textbf{milliseconds} (ms).
The metrics are reported as mean $\pm$ standard deviation (mean $\pm$ std) to reflect both the average performance and its variability.
More detailed analyses are presented in Section \ref{sec:Results}.}

\label{tab: quan}
\setlength{\tabcolsep}{2mm}
\renewcommand{\arraystretch}{1.25}
\scalebox{1.1}{
\begin{tabular}{c|c|ccccc}
\hline
& Methods & PSNR $\!(\uparrow)$ & SAM $\!(\downarrow)$ & RMSE $\!(\downarrow)$ & SSIM $\!(\uparrow)$ & Time (\textbf{ms}) \\
\hline
\multirow{7}{*}{\makecell{Coastline/\\Lake}} 
& MST++ \cite{MST++} & 32.8769$\pm$2.2897 & 4.0060$\pm$0.8073 & 0.0135$\pm$0.0024 & 0.9496$\pm$0.0443 & 15.3939 \\
\cdashline{2-7}
& SPECAT \cite{SPECAT} & 32.1678$\pm$1.8657 & 4.1144$\pm$1.0982 & 0.0141$\pm$0.0033 & 0.9259$\pm$0.0468 & 16.1871 \\
\cdashline{2-7}
& JSSUNet \cite{JSSUNet} & 33.9731$\pm$2.3260 & 3.1173$\pm$0.6129 & 0.0122$\pm$0.0044 & 0.9627$\pm$0.0251 & 46.3122 \\
\cdashline{2-7}
& SSUNet \cite{SSUNet} & 31.8280$\pm$2.7574 & 3.6911$\pm$0.9686 & 0.0140$\pm$0.0037 & 0.9448$\pm$0.0208 & 11.9154 \\
\cdashline{2-7}
& COS2A \cite{COS2A} & 32.6270$\pm$1.9521 & 4.7313$\pm$3.1055 & 0.0157$\pm$0.0066 & 0.9389$\pm$0.0203 & 4.7049 \\
\cdashline{2-7}
& SepSSJSR \cite{SepSSJSR} & 31.8946$\pm$2.0668 & 4.7799$\pm$3.1263 & 0.0165$\pm$0.0064 & 0.9434$\pm$0.0302 & {\bf1.2631} \\
\cdashline{2-7}
& PAINT & {\bf36.2613$\pm$2.2636} & {\bf2.1023$\pm$0.8667} & {\bf0.0101$\pm$0.0042} & {\bf0.9759$\pm$0.0170} & 8.7639 \\
\hline
\multirow{7}{*}{\makecell{Mountain}} 
& MST++ \cite{MST++} & 32.9534$\pm$1.3693 & 2.3733$\pm$0.4238 & 0.0166$\pm$0.0042 & 0.9487$\pm$0.0201 & 15.4946 \\
\cdashline{2-7}
& SPECAT \cite{SPECAT} & 31.9357$\pm$1.4299 & 2.6854$\pm$0.3950 & 0.0184$\pm$0.0047 & 0.9407$\pm$0.0232 & 15.9735 \\
\cdashline{2-7}
& JSSUNet \cite{JSSUNet} & 31.2562$\pm$1.7567 & 2.2163$\pm$0.4417 & 0.0189$\pm$0.0057 & 0.9144$\pm$0.0298 & 46.8671 \\
\cdashline{2-7}
& SSUNet \cite{SSUNet} & 32.0124$\pm$1.7946 & 2.5786$\pm$0.5483 & 0.0186$\pm$0.0056 & 0.9296$\pm$0.0295 & 12.1407 \\
\cdashline{2-7}
& COS2A \cite{COS2A} & 33.2551$\pm$1.7516 & 2.4914$\pm$0.5635 & 0.0185$\pm$0.0058 & 0.9294$\pm$0.0205 & 4.6861 \\
\cdashline{2-7}
& SepSSJSR \cite{SepSSJSR} & 30.4094$\pm$1.6530 & 2.5965$\pm$0.5614 & 0.0215$\pm$0.0061 & 0.9030$\pm$0.0259 & {\bf1.2059} \\
\cdashline{2-7}
& PAINT & {\bf34.3978$\pm$1.4122} & {\bf1.9745$\pm$0.3955} & {\bf0.0152$\pm$0.0046} & {\bf0.9515$\pm$0.0187} & 8.9701 \\
\hline
\multirow{7}{*}{\makecell{Farm}} 
& MST++ \cite{MST++} & 33.2940$\pm$0.7730 & 2.8026$\pm$0.7371 & 0.0163$\pm$0.0014 & 0.9635$\pm$0.0027 & 14.4492 \\
\cdashline{2-7}
& SPECAT \cite{SPECAT} & 32.1656$\pm$0.7917 & 3.1998$\pm$0.7000 & 0.0187$\pm$0.0018 & 0.9527$\pm$0.0052 & 15.3703 \\
\cdashline{2-7}
& JSSUNet \cite{JSSUNet} & 33.7714$\pm$0.7560 & 3.0509$\pm$0.4245 & 0.0145$\pm$0.0014 & 0.9572$\pm$0.0069 & 46.7252 \\
\cdashline{2-7}
& SSUNet \cite{SSUNet} & 32.7604$\pm$0.9700 & 3.0682$\pm$0.7026 & 0.0181$\pm$0.0020 & 0.9502$\pm$0.0043 & 12.1712 \\
\cdashline{2-7}
& COS2A \cite{COS2A} & 33.0971$\pm$0.9576 & 3.1246$\pm$0.8011 & 0.0202$\pm$0.0021 & 0.9402$\pm$0.0058 & 4.9340 \\
\cdashline{2-7}
& SepSSJSR \cite{SepSSJSR} & 31.5462$\pm$0.7604 & 3.2765$\pm$1.0059 & 0.0198$\pm$0.0021 & 0.9379$\pm$0.0053 & {\bf1.2172} \\
\cdashline{2-7}
& PAINT & {\bf35.4473$\pm$0.8148} & {\bf2.1503$\pm$0.4844} & {\bf0.0137$\pm$0.0014} & {\bf0.9678$\pm$0.0035} & 8.1658 \\
\hline
\multirow{7}{*}{\makecell{City}} 
& MST++ \cite{MST++} & 33.7893$\pm$1.4930 & 3.7361$\pm$0.6009 & 0.0157$\pm$0.0047 & 0.9353$\pm$0.0228 & 15.4029 \\
\cdashline{2-7}
& SPECAT \cite{SPECAT} & 33.1386$\pm$1.3758 & 4.1579$\pm$0.4971 & 0.0164$\pm$0.0045 & 0.9287$\pm$0.0225 & 15.5985 \\
\cdashline{2-7}
& JSSUNet \cite{JSSUNet} & 31.7391$\pm$1.4709 & 3.5014$\pm$0.6677 & 0.0174$\pm$0.0050 & 0.9066$\pm$0.0305 & 45.8503 \\
\cdashline{2-7}
& SSUNet \cite{SSUNet} & 33.0905$\pm$1.5376 & 3.9811$\pm$0.5646 & 0.0166$\pm$0.0051 & 0.9237$\pm$0.0262 & 12.3124 \\
\cdashline{2-7}
& COS2A \cite{COS2A} & 33.4620$\pm$1.7359 & 3.9761$\pm$0.6115 & 0.0181$\pm$0.0057 & 0.9139$\pm$0.0255 & 4.7833 \\
\cdashline{2-7}
& SepSSJSR \cite{SepSSJSR} & 30.7675$\pm$1.4246 & 4.2510$\pm$0.7225 & 0.0199$\pm$0.0054 & 0.8944$\pm$0.0265 & {\bf1.2270} \\
\cdashline{2-7}
& PAINT & {\bf34.5762$\pm$1.6477} & {\bf3.0560$\pm$0.4973} & {\bf0.0142$\pm$0.0046} & {\bf0.9401$\pm$0.0226} & 8.8657 \\
\hline
\hline
\multirow{7}{*}{\makecell{Average}} 
& MST++ \cite{MST++} & 33.2284$\pm$1.5581 & 3.2295$\pm$0.9250 & 0.0155$\pm$0.0035 & 0.9492$\pm$0.0299 & 15.1851 \\
\cdashline{2-7}
& SPECAT \cite{SPECAT} & 32.3519$\pm$1.4921 & 3.5794$\pm$0.9824 & 0.0169$\pm$0.0042 & 0.9424$\pm$0.0320 & 15.7823 \\
\cdashline{2-7}
& JSSUNet \cite{JSSUNet} & 32.6849$\pm$2.0181 & 2.9714$\pm$0.8015 & 0.0158$\pm$0.0050 & 0.9352$\pm$0.0349 & 46.4387 \\
\cdashline{2-7}
& SSUNet \cite{SSUNet} & 32.4228$\pm$1.8809 & 3.3298$\pm$0.8819 & 0.0168$\pm$0.0045 & 0.9371$\pm$0.0242 & 12.1349 \\
\cdashline{2-7}
& COS2A \cite{COS2A} & 33.1104$\pm$1.5859 & 3.5808$\pm$1.8269 & 0.0181$\pm$0.0054 & 0.9306$\pm$0.0225 & 4.7770 \\
\cdashline{2-7}
& SepSSJSR \cite{SepSSJSR} & 31.1544$\pm$1.6712 & 3.7260$\pm$1.9154 & 0.0194$\pm$0.0055 & 0.9197$\pm$0.0334 & {\bf1.2283} \\
\cdashline{2-7}
& PAINT & {\bf35.1706$\pm$1.7294} & {\bf2.3209$\pm$0.7139} & {\bf0.0133$\pm$0.0042} & {\bf0.9588$\pm$0.0216} & 8.6913 \\
\hline
\end{tabular}}
\end{table*}

\begin{figure}[t]
\centering
\includegraphics[width=1\linewidth]{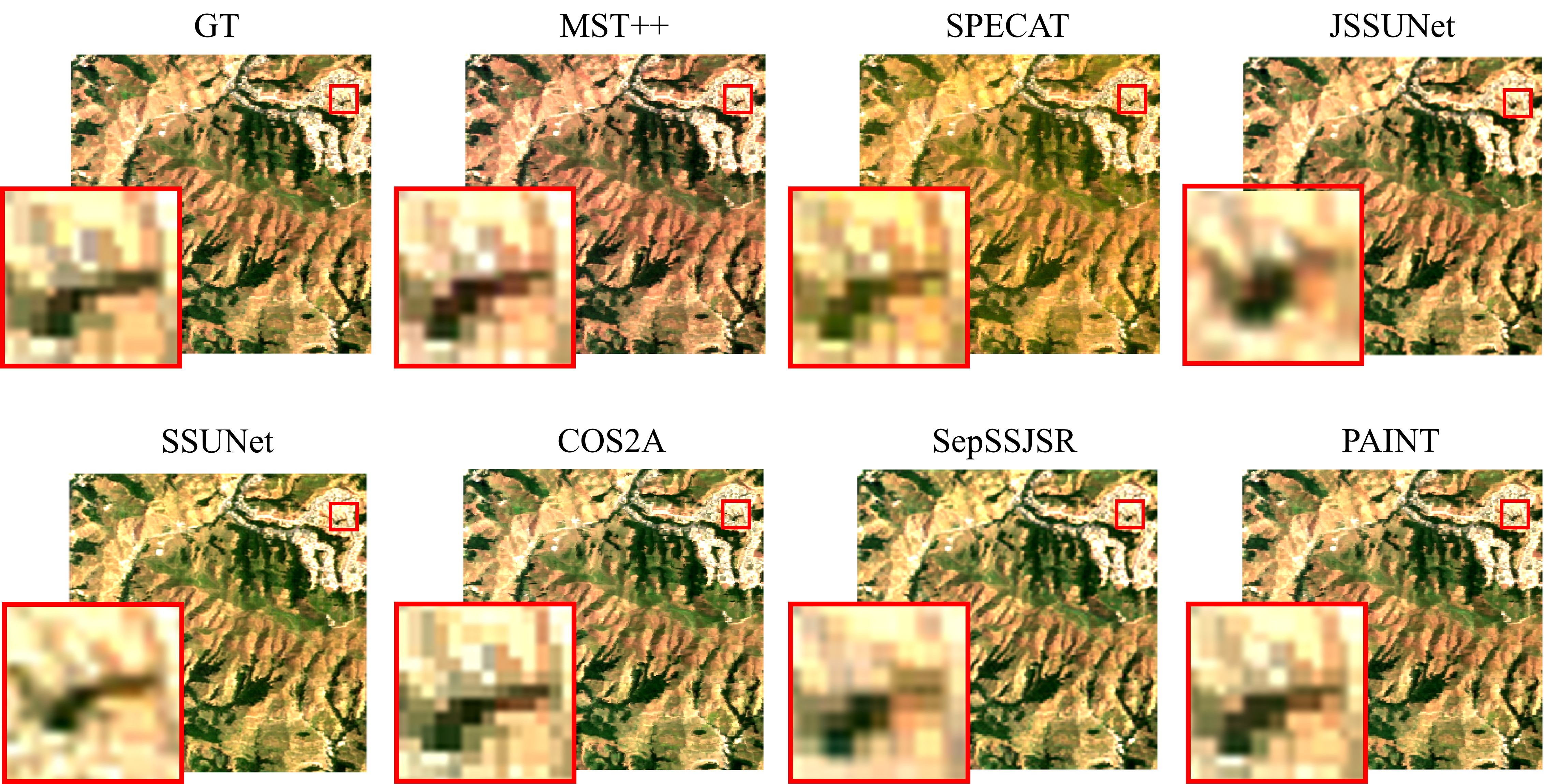}
\caption{Qualitative assessment of spatial fidelity under the mountain scenarios, which contain fine spatial textures.
From left to right, the first row represents the reference HSI and the reconstructed HSIs obtained by MST++, SPECAT, and JSSUNet, respectively.
In addition, the second row illustrates the reconstructed HSIs obtained by SSUNet, COS2A, SepSSJSR, and the proposed PAINT, respectively.
The zoomed-in views of the regions highlighted by small red boxes are further provided to facilitate a better assessment of spatial fidelity.
The qualitative evaluations for the challenging city scenario are provided in Supplementary Figure 2 due to space limitations.
%
%
}\label{fig:spa_compare}
\vspace{-0.3cm}
\end{figure}

We first present the qualitative evaluations of spectral fidelity across the four representative landscapes shown in Figure \ref{fig:spec_curve}.
Furthermore, qualitative evaluations are conducted to examine the reconstructed spatial details, as displayed in Figure \ref{fig:spa_compare}.
Due to space limitations, only a subset of the experimental results is presented in this main article; more detailed results are shown in Supplementary Figure 1 (for spectral fidelity) and Supplementary Figure 2 (for spatial comparison).
In the spectral fidelity evaluation, four pixels are uniformly selected from each landscape for detailed analysis. 
In each subplot corresponding to a selected pixel, the vertical and horizontal axes indicate the reflectance and wavelength range, respectively, while their corresponding spatial locations are marked by red dots, as shown in the true-color compositions on the left of each landscape.
The blue pluses and black curves denote the pixels of input Landsat MSI and their corresponding reference AVIRIS counterparts, respectively. 
Meanwhile, the mint green, orange, brown, yellow, green, purple, and red curves, respectively, indicate the hyperspectral pixels reconstructed using MST++, SPECAT, JSSUNet, SSUNet, COS2A, SepSSJSR, and PAINT.
%

As shown in Figure \ref{fig:spec_curve}(a), the proposed PAINT provides more effective reconstructions across all selected pixels.
The PAINT-reconstructed pixels (i.e., the red curves) closely match the reference pixels (i.e., the black curves) in both shape and reflectance level, yielding reliable reconstructions for the coastline/lake landscape.
In contrast, the peer methods frequently yield noisy results [cf Figure \ref{fig:spec_curve}(a)], which violate the inherent spectral continuity nature of typical hyperspectral data. 
In particular, the spectral curves reconstructed by peer methods at the pixel location (227,11) exhibit pronounced discrepancies relative to the reference curve [cf Figure \ref{fig:spec_curve}(a)], resulting in unsatisfactory performance.
We further notice that, at the location of (39,158) [cf. Figure \ref{fig:spec_curve}(a)], the PAINT-reconstructed pixel and those baseline-reconstructed pixels exhibit a larger reflectance discrepancy in the visible wavelength range; the former is significantly closer to the reference pixels, while the latter are consistent with the Landsat input.
After a careful evaluation, this phenomenon is likely attributed to different capabilities of SpaSR.
Specifically, recovering the HR details from the LR observation inevitably encounters ill-posedness.
Therefore, the algorithms that insufficiently explore HR spatial details likely tend to replicate each LR pixel four times (i.e., $2\times2$ block), which leads to reflectance consistent with the input LR pixel, rather than deriving the true HR details from the observation.
Conversely, our PAINT better reconstructs the spatial details and preserves effective spectral fidelity, yielding an accurate conversion for this landscape. 

For the mountain landscapes [see Fig. \ref{fig:spec_curve}(b)], although the benchmark methods achieve promising spectral reconstructions, the proposed PAINT consistently demonstrates higher spectral fidelity.
As presented in Figure \ref{fig:spec_curve}(b), the hyperspectral pixels reconstructed by benchmark methods still exhibit slight deviations from the reference pixels compared to the proposed PAINT.
In addition, at spatial locations of (15,205) and (243,176), the spectral curves reconstructed by SepSSJSR exhibit more substantial deviations from the reference curve, suggesting limited capability to preserve spectral consistency.
To further evaluate the spatial fidelity, the zoomed-in views of the mountain landscape are provided in Figure \ref{fig:spa_compare}.
These comparisons indicate that the benchmark methods are more susceptible to the blurring effect, suggesting their limited capability to preserve high-frequency spatial details during the SpaSR process. 
This observation is consistent with the results obtained for the coastline/lake landscape.
In addition, the reconstructions produced by MST++ exhibit slight color distortions, indicating its limited ability in such a challenging DualSR task.
In contrast, our PAINT effectively integrates the PDG-Net to recover the HR Landsat MSI, followed by recovering the {\blue target} AVIRIS HSI using ADMM-Net, resulting in a substantially reliable reconstruction.

For the farm and city landscapes [see Figure \ref{fig:spec_curve}(c) and Figure \ref{fig:spec_curve}(d)], the benchmark methods show significant noisy spectral responses and noticeable deviations from the reference curves, especially at pixel locations of (133,135), (47,25), and (204,147), leading to inferior spectral reconstruction performance.
On the other hand, although the signatures reconstructed by the proposed PAINT at pixel locations (25,194) and (171,120) present slight deviations from the reference hyperspectral spectral curve, they still preserve similar reflectance levels and spectral shapes, indicating reliable reconstruction performance.
Therefore, the proposed PAINT achieves superior performance compared to MST++, SPECAT, JSSUNet, SSUNet, COS2A, and SepSSJSR, providing more effective hyperspectral reconstruction for practical applications, such as material identification.

To quantitatively assess the reconstruction performance, the quantitative comparisons are summarized in Table \ref{tab: quan}.
In this comparison, four widely used quantitative metrics are employed, including peak signal-to-noise ratio (PSNR) \cite{CODE}, SAM \cite{SISHY}, root-mean-squared error (RMSE) \cite{HyperQUEEN}, and structural similarity index measure (SSIM) \cite{COCNMF}.
All quantitative results reported in Table \ref{tab: quan} are averaged over all samples within the testing dataset.
As demonstrated in the table, the proposed PAINT substantially outperforms the baselines in all metrics across all landscapes.
Specifically, in scenarios characterized by homogeneous materials (e.g., coastline/lake and farm), PAINT demonstrates high-quality reconstruction performance, with PSNR values exceeding 35 dB in both scenarios.
Even in challenging landscapes characterized by heterogeneous materials (i.e., mountain and city), PAINT maintains excellent performance, achieving PSNR values above 34.3 dB and outperforming the runner-up benchmark method (i.e., MST++) {\blue by approximately 1 dB}.
The potential reason is that the transformer-based methods (e.g., MST++ and SPECAT) rely on computationally expensive attention mechanisms and lack interpretability, which may hinder their network optimization and consequently result in unsatisfactory solutions. 
On the other hand, COS2A was originally developed for AVIRIS-level hyperspectral reconstruction from 12-band MSI, which is theoretically less challenging than our target 7-to-172 DualSR task, thereby resulting in degraded performance when adapted to this problem.
Although JSSUNet adopts a deep unfolding framework, its reconstruction primarily relies on learnable cluster-specific operators without explicitly enforcing spectral continuity, thereby limiting its capability for the 7-to-172 DualSR task.
SSU-Net effectively exploits spectral and spatial features through a dual-branch U-Net architecture. 
However, it remains a purely data-driven framework and lacks physical constraints, making it unable to address the highly ill-posed 7-to-172 DualSR task.
On the other hand, SepSSJSR treats SpaSR and SpeSR as separate tasks and trains the networks independently.
Moreover, these networks comprise only non-interpretable CNN architectures, which may limit their ability to capture reliable global features.
These limitations potentially render SepSSJSR a suboptimal solution.
Overall, the proposed PAINT yields an average improvement of nearly 2 dB, 2.8 dB, 2.5 dB, 2.8 dB, 2 dB, and 4 dB in PSNR over MST++, SPECAT, JSSUNet, SSUNet, COS2A, and SepSSJSR, respectively.
Furthermore, PAINT also improves the SAM metrics by around 1 degree compared to the benchmark methods.
Consistent performance improvements can also be observed in terms of RMSE and SSIM, as presented in Table \ref{tab: quan}. 
In addition, PAINT not only substantially improves the average reconstruction performance while maintaining comparable variability, but also performs DualSR in near real-time with millisecond-level inference speeds. 
These advantages demonstrate the reliability of the reconstructed spectral signatures across diverse landscapes and their practical applicability to downstream RS and EO tasks.
Consequently, the above qualitative and quantitative analyses have demonstrated the effectiveness and superiority of the proposed PAINT compared to the baselines.
In the following experiments, we conduct case studies to demonstrate the practical utility of our framework.
\subsection{Case Study: Classification Evaluation}\label{sec: Classification}
\begin{figure}[t]
\centering
\includegraphics[width=1.0\linewidth]{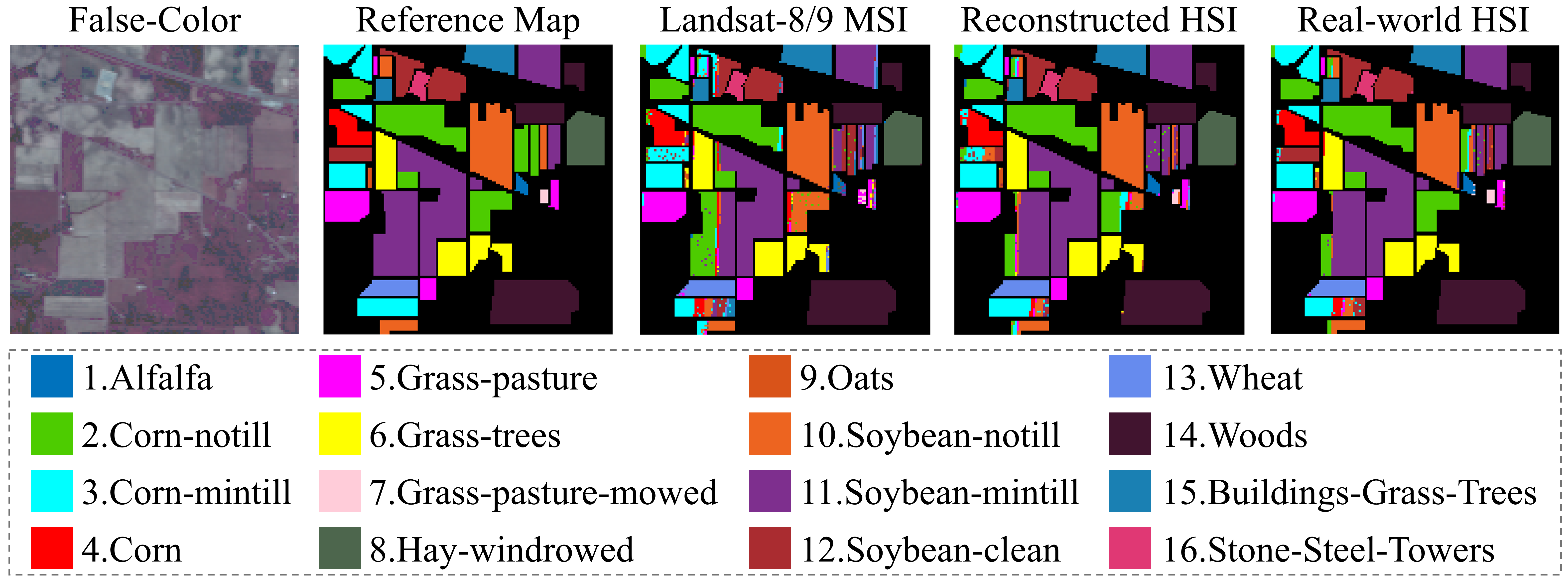}
\caption{Visualizations of the false-color composition, reference map, and classification results with various input representations on the Indian Pines dataset.
Reconstructed HSI (returned by our PAINT) does yield much stronger classification results compared to the Landsat MSI.
The color blocks and indices denote the corresponding classes, while the quantitative performance is summarized in Table \ref{tab: classification}.}\label{fig:classification}
\vspace{-0.3cm}
\end{figure}

In this section, we further conduct a case study on the pixel-level classification task to demonstrate the practicality of the proposed PAINT framework.
Specifically, HR HSIs that contain hundreds of spectral bands (e.g., AVIRIS) provide substantially richer spectral information than LR MSIs with a limited number of spectral bands (e.g., Landsat-8/9).
Therefore, HR hyperspectral data {\blue are} especially suitable for exploiting material identification applications.
{\blue For instance, the popular research topic of pixel-wise HSI classification has attracted considerable attention in the recent RS community \cite{gao2023lica,tang2025hypereast,huang2024adaptive}.}
%
%
Accordingly, in the following experiments, we evaluate the practical applicability of our PAINT framework by using Landsat-level MSI, PAINT-reconstructed HSI, and real-world HSI as inputs to a hyperspectral classification (HC) algorithm, and then compare their classification performance.
A reliable reconstruction is expected to significantly improve the classification performance of Landsat-level MSI and also achieve competitive results with those from real-world HSIs.
In these experiments, the superpixelwise principal component analysis (SuperPCA) algorithm \cite{SuperPCA} is adopted as the classifier.
The SuperPCA framework first performs unsupervised feature extraction by exploiting the input data structure, and subsequently applies the support vector machine (SVM) to classify the extracted features.
Consequently, a high-quality reconstruction is required to precisely preserve realistic hyperspectral structures in order to achieve an effective classification performance. 
The experimental settings of SuperPCA classification are summarized in Supplementary Note 4 due to space limitations.
On the other hand, the representative Indian Pines dataset (cf. Figure \ref{fig:classification}), which was captured by the AVIRIS sensor, is employed.
The official Indian Pines dataset consists of $L=145\times 145$ pixels with 200 spectral bands after removing bands affected by water absorption, while the labeled pixels include a total of 16 classes (cf. Figure \ref{fig:classification}).
Additional details of the Indian Pines dataset can be found in \cite{IPdata}.
Due to the lack of spatially and temporally aligned Landsat-8/9 data for the Indian Pines dataset, we simulate the input Landsat-level MSI (resp. panchromatic image) to enable this case study.
The corresponding details are also presented in Supplementary Note 4.
Given the Landsat-level MSI and panchromatic image, the corresponding PAINT-reconstructed HSI for this Indian Pines data can be obtained.
Consequently, the effectiveness of the proposed PAINT is evaluated by comparing classification performance across various input representations.
%

\begin{table}[t]
\centering\scriptsize
\caption{Comparisons of classification performance using various input representations of the Indian Pines data, including Landsat-level MSI, PAINT-reconstructed HSI, and the real HSI.
Detailed analyses are provided in Section \ref{sec: Classification}.
}
\label{tab: classification}
\begin{tabular}{c|c|c|c}
     \hline\hline
     Class No. & Landsat-level MSI & Reconstructed HSI  & Real-world HSI  \\
     
     \hline
     1  & 56.09$\pm$34.14 & 92.61$\pm$7.11 & 96.52$\pm$5.34 
     
     \\
     
     2  & 59.82$\pm$8.68 & 81.62$\pm$4.10 & 88.79$\pm$2.18 
     
     \\
     
     3  & 67.22$\pm$10.58 & 85.28$\pm$4.16 & 84.25$\pm$3.10 
      
     \\
     
      4  & 83.65$\pm$5.53 & 91.08$\pm$4.14 & 91.86$\pm$2.33  
      
     \\
     
      5 & 89.69$\pm$5.67 & 96.20$\pm$2.10 & 97.99$\pm$1.79  
      
      \\

     6  & 93.79$\pm$1.09 & 98.89$\pm$0.58 & 98.03$\pm$1.25 
      
      \\

      7 & 35.00$\pm$38.75 & 57.86$\pm$26.18 & 81.43$\pm$21.35
      
      \\

     8  & 96.96$\pm$2.39 &  99.80$\pm$0.10 & 100.00$\pm$0.00 
      
      \\

      9  & 6.00$\pm$8.43 & 95.00$\pm$7.07 & 84.00$\pm$16.47  

       \\

      10  & 84.77$\pm$3.67 & 89.97$\pm$1.83 & 87.44$\pm$1.66 

       \\

      11 & 72.01$\pm$8.65 & 93.44$\pm$2.18 & 89.27$\pm$4.15  

       \\

      12 & 63.31$\pm$12.53 & 87.86$\pm$5.19 & 91.70$\pm$3.20 

       \\

      13  & 96.30$\pm$1.35 & 99.63$\pm$0.39 & 99.26$\pm$0.00 
       \\

      14  & 99.62$\pm$0.68 & 99.60$\pm$0.28 & 100.00$\pm$0.00   

         \\

      15 & 98.61$\pm$0.99 & 98.64$\pm$1.09 & 99.46$\pm$0.65 

         \\

      16  & 99.57$\pm$1.37 & 95.65$\pm$3.55 & 96.96$\pm$2.10
      \\
      \hline
     {OA(\%)}& 78.98$\pm$1.63 & 92.16$\pm$1.06 & 92.18$\pm$1.28
     \\
     {AA(\%)}& 75.15$\pm$2.81 & 91.45$\pm$2.30 & 92.94$\pm$0.91  
     \\
     {$\kappa$($\times 100$)} & 76.06$\pm$1.79 & 90.98$\pm$1.22 & 91.01$\pm$1.46
     \\

    \hline\hline
\end{tabular}
\end{table}
%
The classification maps obtained by using the Landsat-level MSI, PAINT-reconstructed HSI, and the real-world HSI are illustrated in Figure \ref{fig:classification}.
To facilitate a fair evaluation, the quantitative metrics of overall accuracy (OA), averaged accuracy (AA), and the Kappa coefficient ($\mathcal{K}$) are reported in Table \ref{tab: classification}.
Notably, the reported metrics are averaged over ten independent runs to prevent randomness from affecting the performance comparisons.
As demonstrated in these results, the proposed PAINT significantly upgrades the Landsat classification performance of 78.98\% OA and 76.06\% Kappa to 92.16\% OA and 90.98\% Kappa, which is competitive with the real-world HSI (92.18\% OA and 91.01\% Kappa).
Due to space limitations, we further present an in-depth discussion of the classification results in Supplementary Note 5.
In summary, this case study on classification strongly supports the practical applicability of the PAINT framework.

\subsection{Case Study: BSS Evaluation}\label{sec: BSS}
\begin{figure}[t]
\centering
\includegraphics[width=0.9\linewidth]{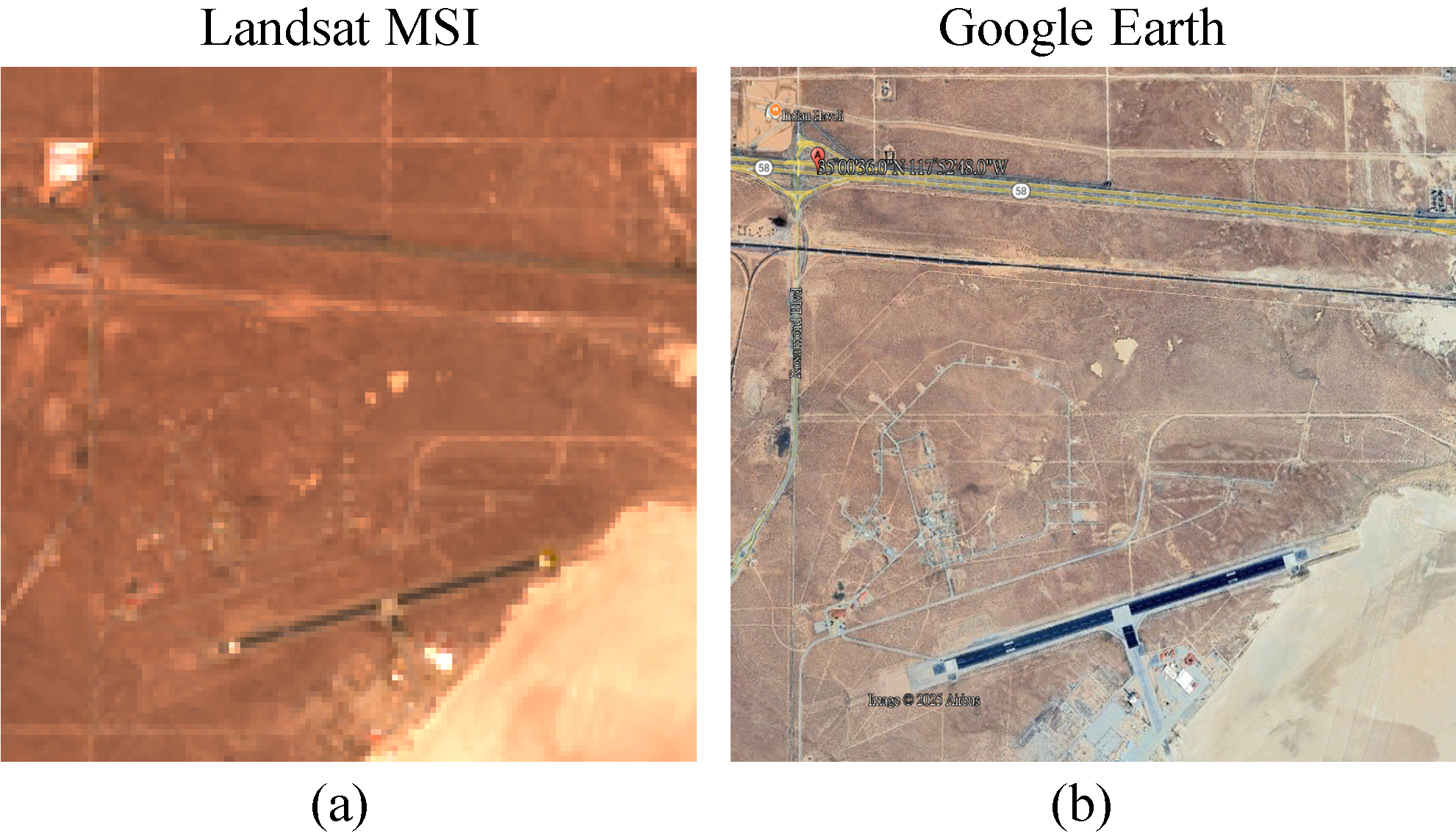}
\caption{(a) True-color composition of the ROI observed by Landsat-8/9 satellite.
(b) HR view of the ROI obtained by Google Earth imagery.
The ROI is located near Edwards Air Force North Base in the Mojave Desert, California, USA, and mainly consists of desert, dry lakebed, runway, and buildings.
The corresponding BSS results are illustrated in Figure \ref{fig:BSS_result}.
}\label{fig:MDL}
\vspace{-0.4cm}
\end{figure}

Beyond the case study on the HC task, another fundamental signal processing task for material understanding and analysis is blind source separation (BSS).
Given a $B$-band remotely sensed image $\bX\in\mathbb{R}^{B\times L}$, BSS algorithms aim to recover $N\ll B$ pure spectral endmembers $\bE\in\mathbb{R}^{B\times N}$ and their corresponding spatial abundance maps $\bA\in\mathbb{R}^{N\times L}$ such that $\bX\approx\bE\bA$ \cite{8633850,MU2026TIP,HyperCSI}.
However, accurate BSS heavily relies on the rich spectral information, as provided by AVIRIS-level HSIs, to enable the reliable recovery of both endmembers and their abundance maps.
When the number of spectral bands is limited, as in Landsat-8/9 MSIs, BSS often suffers from degraded recovery, {\blue such as} repeated endmembers with misassigned abundance maps \cite{PRIME,MU2026TIP}.
Since the proposed PAINT framework is designed to convert the Landsat-8/9 MSIs into their AVIRIS-level counterparts, the reconstructed HSIs are expected to significantly improve the BSS performance.
Accordingly, this case study further demonstrates that the PAINT-reconstructed HSI can facilitate superior BSS results compared to those obtained by using real-world Landsat-8/9 MSI (cf. Figure \ref{fig:MDL}),  and the experiments are detailed below.
To further demonstrate this effectiveness, a real-world Landsat-8/9 MSI, as shown in Figure \ref{fig:MDL}(a), is employed for this case study.
This ROI was acquired on May 16, 2024, over the  Mojave Desert, California, USA, near Edwards Air Force North Base. 
The LR Landsat-8/9 MSI $\bY_L$ contains $L=128\times 128$ pixels and consists mainly of desert, dry lakebed, runway, and buildings.
Since the reference endmembers and the corresponding abundance maps are typically unavailable for real-world Landsat-8/9 BSS, the HR view of the same ROI obtained from Google Earth imagery [cf. Figure \ref{fig:MDL}(b)] is additionally incorporated to support material interpretation and understanding.
According to the Google Earth imagery, the high-brightness objects located at the upper-left and lower-right corners correspond to buildings.
The first one is likely to be an Indian-Hawill restaurant, while the latter one {\blue appears to be} Edwards Air Force Base.
Moreover, the diagonal line and two nearly parallel horizontal lines correspond to the Air Force runway and the Barstow-Bakersfield highway, respectively, which are likely composed of similar materials.
In the background, this ROI predominantly comprises desert, appearing as a spatially extensive dark-brown region of Landsat MSI, together with a dry lakebed that is shown in the triangular area located in the lower-right corner (cf. Figure \ref{fig:MDL}).
%

%
In this case study, the hyperplane-based Craig-simplex-identification (HyperCSI) \cite{HyperCSI} is employed for BSS.
HyperCSI has been theoretically proven to recover the true endmembers given a sufficient number of pixels \cite[Theorem 2]{HyperCSI}, without relying on the less practical pure-pixel assumption \cite{SPA}.
The model order $N$ is further determined using the information-theoretic, unsupervised model order selection algorithm, the minimum description length (MDL) criterion \cite{MDL}, applied to the reconstructed HSI, because the insufficient spectral dimension of Landsat-8/9 MSI limits code length estimation as $N$ increases.
The minimum coding length is achieved when $N:=4$, aligning with the observation from the Google Earth imagery.
In HyperCSI, the radius and simplex compression ratios are empirically set to $10^{-8}$ and 0.99, respectively.
With the above settings, we proceed to the comparisons of the real-world BSS results obtained by Landsat-8/9 MSI and its PAINT-reconstructed HSI.
The qualitative comparisons of the BSS results are presented in Figure \ref{fig:BSS_result}, while the detailed analyses are presented collectively in Supplementary Note 6 due to limited space.
The BSS results again demonstrate the effectiveness and practical utility of our PAINT framework.
Thus far, comprehensive experiments have been conducted, encompassing both qualitative and quantitative evaluation, pixel-level classification, and this BSS case study.
All results substantially support the proposed PAINT algorithm.

\begin{figure}[t]
\centering
\includegraphics[width=1.0\linewidth]{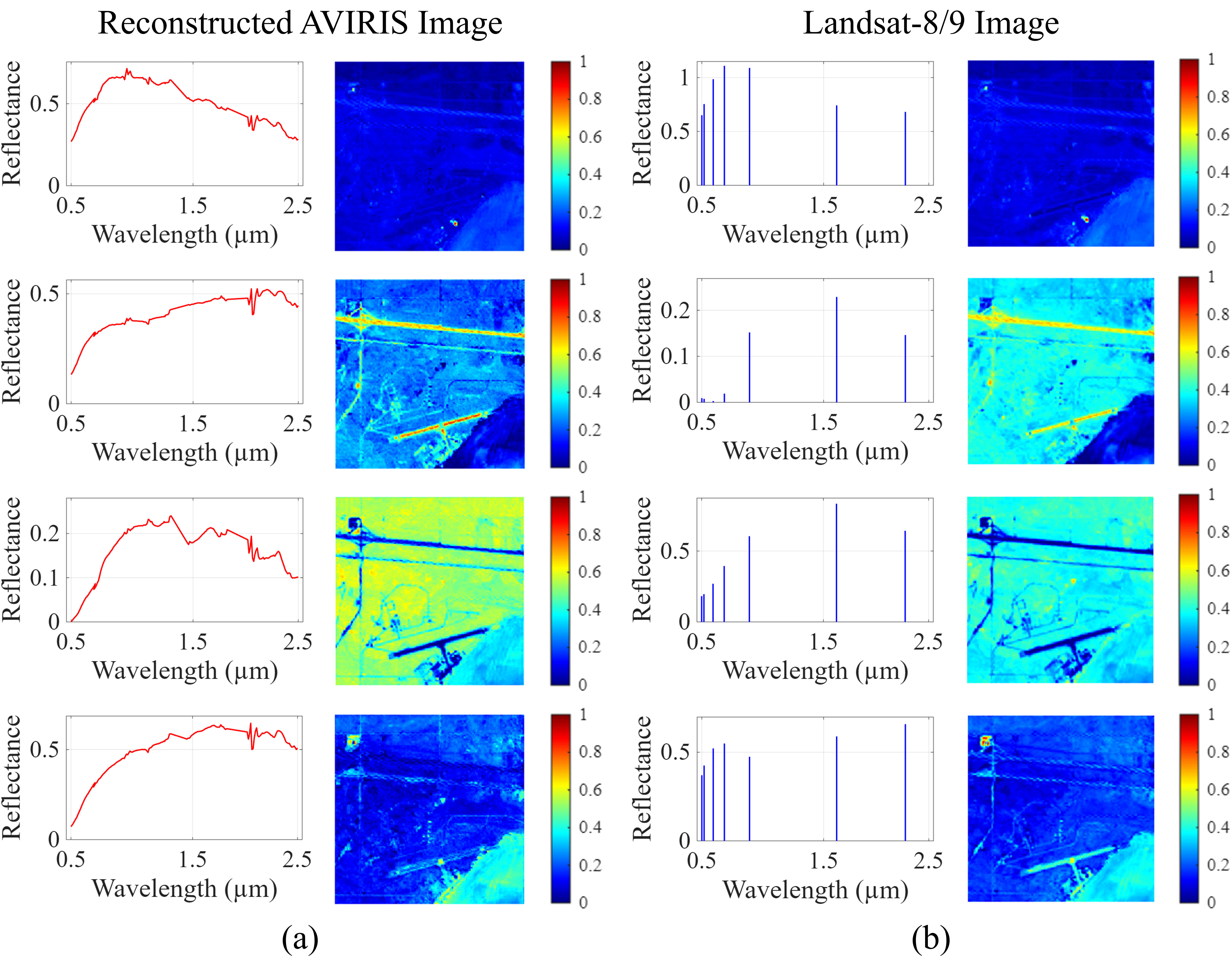}
\caption{(a) Hyperspectral endmembers (i.e., the red curves) and the corresponding abundance maps recovered from the PAINT-reconstructed HSI using HyperCSI BSS \cite{HyperCSI}.
(b) Multispectral endmembers (i.e., the blue pulses) and the corresponding abundance maps recovered from the Landsat-8/9 MSI using HyperCSI BSS.
The model order $N:=4$ is rigorously determined by the MDL criterion \cite{MDL}.
Further analyses are summarized in Section \ref{sec: BSS}.
}\label{fig:BSS_result}
\end{figure}

\subsection{Further Discussion}\label{sec: Discussion}
\subsubsection{Spectral Hallucination and Reconstruction Reliability}
{In real-world applications, faithfully recovering the underlying spectral information is far more important than generating spectral signatures that appear to be plausible yet reflect inaccurate characteristics \cite{NEURIPS2024_2847d43f}.
To provide insights for future research, we further analyze the reconstruction reliability, especially highlighting which spectral information can be reliably recovered.
As shown in Figure \ref{fig:spec_curve} and Figure \ref{fig:BSS_result}(b), the Landsat MSI provides only three spectral bands in the wavelength range of 0.8--2.5 $\mu$m, rendering the reconstruction of the corresponding hyperspectral signatures highly ill-posed and increasing the risk of hallucinated spectral information.
For example, as shown in Figure \ref{fig:spec_curve}(a), the benchmark methods can reasonably reconstruct the spectral signatures within the wavelength range of 0.5--0.8~$\mu$m, where spectral observations (i.e., the four leftmost blue impulses) from Landsat MSI are relatively dense.
In contrast, within the wavelength range of 0.8--2.5~$\mu$m (corresponding to the three rightmost blue impulses), the spectral curves reconstructed by most baselines exhibit noticeable deviations from the reference HSIs, e.g., at the spatial location of (227, 11) over the coastline landscape.
This observation suggests that the limited spectral observations in this wavelength range increase the risk of hallucinated spectral information.

A similar phenomenon can be observed in the other representative landscapes, particularly at the spatial locations (243, 176), (133, 135), and (171, 120) for the mountain, farm, and city landscapes, respectively.
Consequently, interpretability becomes especially important when addressing such a challenging Dual task. 
In contrast to these baseline methods, PAINT achieves more reliable spectral reconstruction across the wavelength range of 0.8–2.5 $\mu$m, which can be attributed to its theoretically grounded and highly interpretable designs (cf. Section \ref{sec:thm}).
Overall, the experiments show that spectral information within 0.5–0.8 $\mu$m can be more
faithfully recovered by most methods due to the denser Landsat observations. 
Reconstructed bands within wavelengths of 0.8–2.5 $\mu$m are substantially more uncertain and may contain underlying spectral hallucinations, thereby requiring more reliable reconstruction, such as the proposed PAINT.

\subsubsection{Robustness to Data Simulation Protocols}
In this section, we conduct an additional sensitivity analysis of the adopted data simulation protocol.
Specifically, due to the lack of a publicly available spectral response function (SRF) for characterizing AVIRIS HSI bands to the corresponding Landsat-8/9 MSI bands, we follow the widely recognized and frequently adopted band-averaging strategy \cite{SepSSJSR}, which corresponds to a uniform SRF.
Although models trained on such synthetic data pairs have demonstrated promising practical applicability to downstream RS tasks (cf. Sections \ref{sec: Classification} and \ref{sec: BSS}), the potential bias may be introduced by this simulation protocol and hence requires careful consideration.
To provide a more comprehensive evaluation, we conducted an additional sensitivity analysis by directly comparing the simulated Landsat-8/9 MSI generated using different simulation protocols with the corresponding real Landsat-8/9 MSI (cf. Table \ref{tab:data_combined}).

Specifically, as shown in Figure \ref{fig:similarity_map}, we compare the simulation quality of the band-averaging strategy (i.e., uniform SRF) with that of the commonly used Gaussian and triangular SRF \cite{moradi2025generating} using cosine similarity error maps \cite{zhao2023few}.
This metric effectively reflects the spectral similarity between the simulated and real Landsat-8/9 MSI.
Moreover, the cosine similarity error maps are computed by subtracting the pixel-wise cosine similarity scores from 1, thereby resulting in a metric where a lower value indicates better simulation quality. 
The results demonstrate that the adopted uniform SRF provides a close approximation to the real Landsat-8/9 observations and achieves simulation quality comparable to that of the widely used Gaussian and triangular SRFs. 
This observation indicates that the simulation
results are largely insensitive to the specific shape of the adopted SRF. 
Together with the comprehensive experiments (cf. Section \ref{sec: experiment}), these findings support the use of the uniform SRF as a sufficiently effective approximation for synthesizing Landsat-8/9 MSI.
\begin{figure}[t]
\centering
\includegraphics[width=1.0\linewidth]{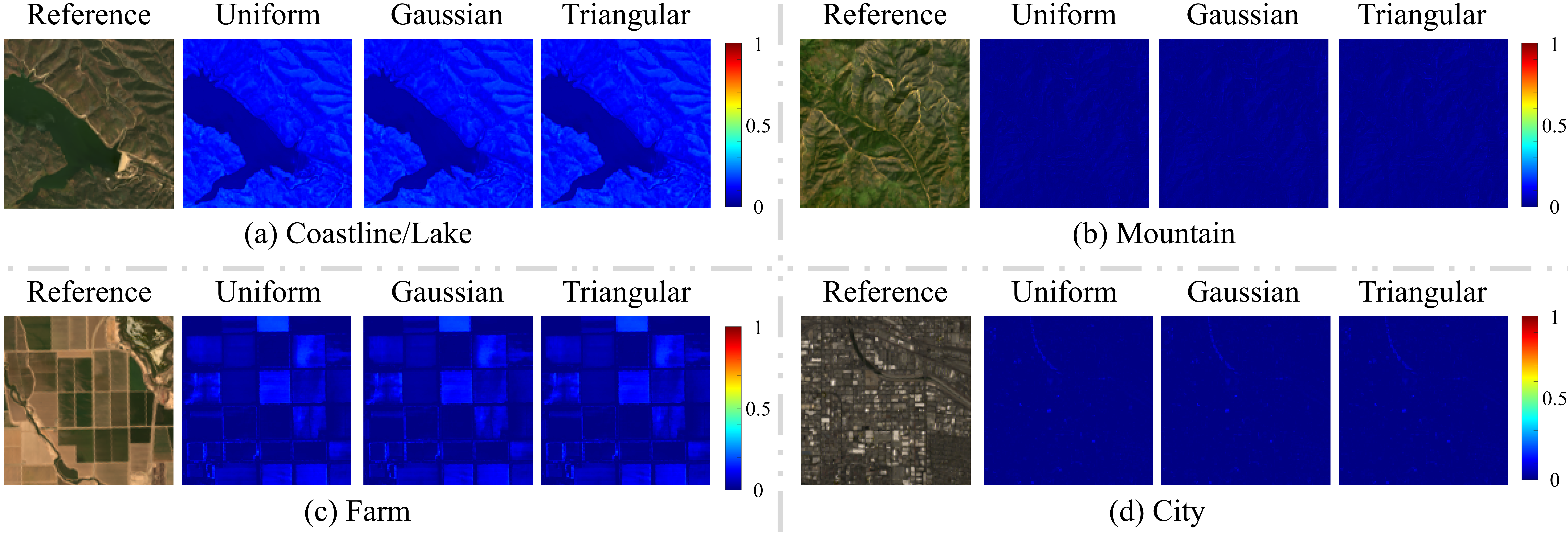}
\caption{To validate the proposed data simulation protocol, true-color RGB images of the real Landsat-8/9 MSI and the corresponding cosine similarity error maps obtained using the uniform spectral response function (SRF), Gaussian SRF, and triangular SRF are presented across four representative landscapes: (a) coastline/lake, (b) mountain, (c) farm, and (d) city. 
The cosine similarity error is computed by subtracting cosine similarity from 1, where lower values indicate higher spectral similarity.
}\label{fig:similarity_map}
\end{figure}

}

\subsection{Ablation Study}\label{sec:ablation} 
{
To provide a deeper understanding of the effectiveness of the proposed ADMM-Net, PGD-Net, spectral continuity prior module (SCPM), as well as the necessity of the panchromatic image, we conducted an ablation study, with the results summarized in Table \ref{tab:ablation}.
In addition, we have also conducted an in-depth experiment to evaluate the learnable Woodbury lemma module (cf. Model 2 in Figure \ref{fig:network})
%
%
The effectiveness of each component is discussed in detail below.
\begin{table}[t]
\caption{Ablation study of the proposed PAINT algorithm to evaluate the effectiveness of the PGD-Net, the spectral continuity prior module (SCPM), and the input panchromatic image (PAN). 
The first row corresponds to the baseline ADMM-Net (i.e., PAINT with only ADMM-Net), while the fourth row represents the complete PAINT framework.
Further analyses can be found in Section \ref{sec:ablation}.
}
\renewcommand\arraystretch{1.15}
\begin{center}
\setlength{\tabcolsep}{1.25mm}
\scalebox{1}{
\begin{tabular}{c c c|c c c c}
\hline
\hline
\makecell[c]{PGD-Net} & \makecell[c]{SCPM} & \makecell[c]{PAN} & \makecell[c]{PSNR ($\uparrow$)} & \makecell[c]{SAM ($\downarrow$)} & \makecell[c]{RMSE ($\downarrow)$} & \makecell[c]{SSIM ($\uparrow)$}

\\
\hline
\ding{55} & \ding{55} & \ding{55}
& 32.1527 & 2.7540  & 0.0169 & 0.9293
\\

\hline
\ding{55} & \ding{51} & \ding{55}
& 32.9639 & 2.4274  & 0.0155 & 0.9367
\\

\hline
\ding{51} & \ding{55} & \ding{51}
& 32.7094 & 2.9689 & 0.0165 & 0.9400
\\

\hline
\ding{51} & \ding{51} & \ding{51}
& {\bf35.1706} & {\bf2.3209}  & {\bf0.0133} & {\bf0.9588}
\\

\hline
\hline
\end{tabular}
}
\label{tab:ablation}
\end{center}
\end{table}

As shown in Table \ref{tab:ablation}, the first row presents the baseline ADMM-Net (i.e., PAINT with only ADMM-Net), while the remaining configurations are designed to evaluate the contributions of the SCPM, PGD-Net, and the necessity of panchromatic input within the proposed PAINT framework.
%
%
In Table \ref{tab:ablation}, the first-row configuration directly reconstructs the 172-band using merely ADMM-Net, where the 2-time spatially upsampled Landsat MSIs are used as input to ensure the spatial consistency; this ablation case is referred to as the baseline ADMM-Net. 
The second-row configuration additionally incorporates the proposed SCPM with the baseline ADMM-Net.
 The third-row configuration introduces the proposed PGD-Net with panchromatic input to reconstruct an HR Landsat-8/9 MSI, followed by using the baseline ADMM-Net (with  HR Landsat-8/9 MSI as input) to address the 7-172 spectral reconstruction.
 The configuration presented in the fourth row of Table \ref{tab:ablation} refers to the proposed PAINT.

As presented in the first row of Table \ref{tab:ablation}, the baseline ADMM-Net achieves comparable performance to benchmark methods (cf. Table \ref{tab: quan}).
After integrating the SCPM (cf. the second row of Table \ref{tab:ablation}), the reconstruction performance is further improved across all evaluation metrics.
This observation confirms that the spectral continuity prior in HSIs can be effectively exploited for SpeSR with negligible computational cost.
In addition, incorporating PGD-Net and the panchromatic input (cf. the third row of Table \ref{tab:ablation}) improves performance across all evaluation metrics compared to the baseline ADMM-Net.
This result indicates that the proposed PGD-Net effectively exploits the HR spatial information provided by the panchromatic band, thereby benefiting the downstream SpeSR task.
Furthermore, as demonstrated in the third and fourth rows of Table \ref{tab:ablation}, incorporating the lightweight SCPM improves the PSNR by approximately 2.5 dB while consistently enhancing the other evaluation metrics.
These results demonstrate that both the customized PGD-Net and the SCPM individually improve reconstruction quality.
Finally, to validate their complementary effects, we jointly incorporate PGD-Net, SCPM, and panchromatic input to leverage refined spatial information and spectral continuity.
As shown in the first and fourth rows of Table \ref{tab:ablation}, their joint integration consistently achieves the best performance across all evaluation metrics, demonstrating the complementary benefits of spatial refinement, spectral continuity, and HR spatial information for the DualSR task.

On the other hand, to further evaluate the effectiveness of the W-lemma module, we construct a model by removing the SFC layer (cf. Model 2 in Figure \ref{fig:network}) from the baseline ADMM-Net, and then evaluate its performance under the same experimental setting as the baseline ADMM-Net.
The experimental results demonstrate that the PSNR of the baseline ADMM-Net without the W-lemma module degrades by nearly 0.35 dB relative to that of the baseline ADMM-Net, which substitutes the effectiveness of the W-lemma module.
Overall, these results verify the effectiveness of the W-lemma module, customized PGD-Net, the SCPM, and the panchromatic input, highlighting their complementary roles in the proposed PAINT framework.
}

\section{Conclusions and Future Works}\label{sec:conclusion}

To enable reliable global hyperspectral monitoring, this study, for the first time, interpretably converts 30-m 7-band Landsat-8/9 MSI into 15-m 172-band AVIRIS-level HSI via ADMM and PGD optimization theory.
For the highly ill-posed SpeSR problem (7-band to 172-band), we leverage the Woodbury matrix identity and a spectral continuity prior to design the computationally efficient SpeSR algorithm under an ADMM optimization scheme.
Furthermore, we employ the PGD optimization theory to avoid LMI implementations, thereby enabling an efficient and reliable SpaSR (30-m to 15-m) algorithm. 
With these theoretically oriented designs, the challenging Landsat-to-AVIRIS conversion can be elegantly and efficiently addressed.
Otherwise, even for those best existing CAVE-level super-resolution methods (only 31 visible bands), the challenging target mission is hardly achieved.

To facilitate real-time computing for downstream RS and EO tasks, we further derive closed-form expressions using convex optimization theory and, accordingly, deploy interpretable deep unfolding network elements/modules using DIP theory and the physical meanings of the expressions.
We refer to the proposed global hyperspectral monitoring framework as PGD-ADMM interpretable network (PAINT).
As it turns out, PAINT reconstructs high-fidelity spectral shapes (indicated by SAM) and high-precision spatial textures (cf. Figure \ref{fig:spa_compare}), and achieves state-of-the-art DualSR performance with millisecond-level computing.
The ablation study further demonstrates that the superior performance of PAINT stems from the complementary contributions of the proposed ADMM-Net, PGD-Net, and spectral continuity prior.
The integration of these components enables PAINT to effectively exploit both HR spatial information and spectral continuity, consistently achieving the best reconstruction performance across all landscapes.

PAINT has also been shown to be general across diverse landscapes.
More importantly, PAINT-reconstructed HSI yields significantly better material classification and source separation results (cf. Table \ref{tab: classification} and Figure \ref{fig:BSS_result}) than original Landsat MSI, thereby demonstrating the reliability of PAINT for real-world RS and EO applications in the future.
These outstanding material classification and identification abilities of PAINT-reconstructed HSIs, together with the very wide spatiotemporal coverage of NASA's historical Landsat data, should spur numerous novel future RS applications, such as hyperspectral-empowered precision agriculture, real-time forest monitoring, and underground anomaly detection and tracking.
Furthermore, we plan to introduce quantum AI \cite{HyperQUEEN} or quantum generative AI \cite{HyperKING} to have an even preciser spectral feature learning, as quantum entanglement has been recently proven critical in capturing fine details and sharp fluctuations \cite{SquareMamba}, which are essential for SpeSR and DualSR tasks.



\bibliography{ref}

\begin{IEEEbiography}[{\resizebox{0.9in}{!}{\includegraphics[width=1in,height=1.25in,clip,keepaspectratio]{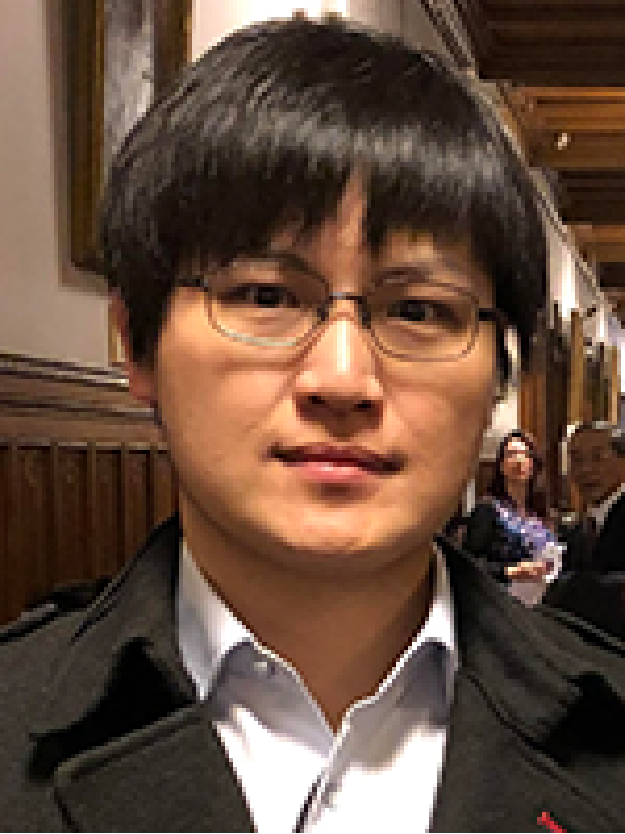}}}]
{\bf Chia-Hsiang Lin}
(S'10-M'18-SM'24)
received the B.S. degree in electrical engineering and the Ph.D. degree in communications engineering from National Tsing Hua University (NTHU), Taiwan, in 2010 and 2016, respectively.
From 2015 to 2016, he was a Visiting Student of Virginia Tech,
Arlington, VA, USA.

He is currently a Professor with the Department of Electrical Engineering,
National Cheng Kung University (NCKU), Taiwan, and also serves as a Technical Director of Smart Sensing \& Systems Technology Center, Industrial Technology Research Institute (ITRI).
Before joining NCKU, he held research positions with The Chinese University of Hong Kong, HK (2014 and 2017),
NTHU (2016-2017),
and the University of Lisbon (ULisboa), Lisbon, Portugal (2017-2018).
He was an Assistant Professor with the Center for Space and Remote Sensing Research, National Central University, Taiwan, in 2018, a Visiting Professor with ULisboa, in 2019, and a Visiting Professor with Texas A\&M University, USA, in 2025.
His research interests include network science,
quantum computing,
convex geometry and optimization, blind signal processing, and imaging science.

Dr. Lin received the Emerging Young Scholar Award (The 2030 Cross-Generation Program) from National Science and Technology Council (NSTC), from 2023 to 2027,
the Future Technology Award from NSTC, in 2022,
the Outstanding Youth Electrical Engineer Award from The Chinese Institute of Electrical Engineering (CIEE), in 2022,
the Best Young Professional Member Award from IEEE Tainan Section, in 2021,
and the Prize Paper Award from IEEE Geoscience and Remote Sensing Society (GRS-S), in 2020.
He received the Ministry of Science and Technology (MOST) Young Scholar Fellowship, together with the EINSTEIN Grant Award, from 2018 to 2023.
In 2016, he was a recipient of the Outstanding Doctoral Dissertation Award from the Chinese Image Processing and Pattern Recognition Society and the Best Doctoral Dissertation Award from the IEEE GRS-S.
\end{IEEEbiography}

\begin{IEEEbiography}[{\resizebox{1in}{!}{\includegraphics[width=1in,height=1.25in,clip,keepaspectratio]{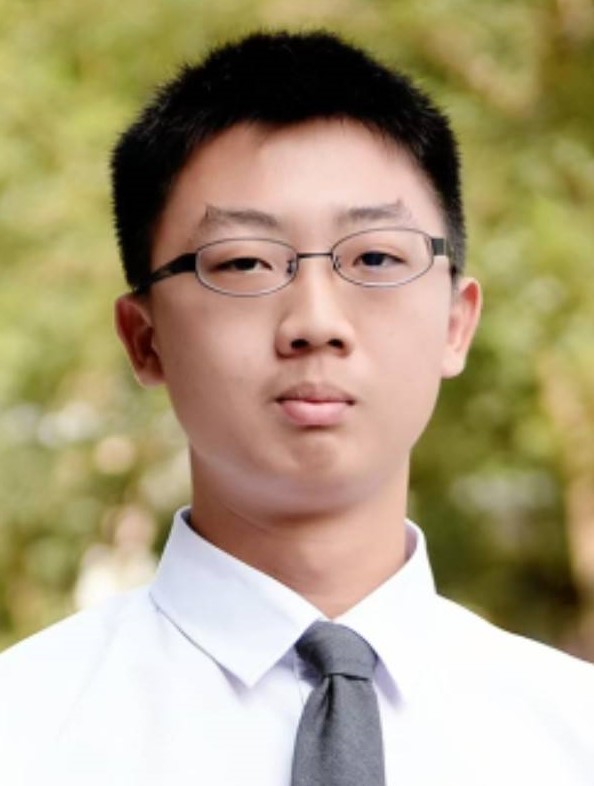}}}]
{\bf Jian-Kai Huang} (S'25) received the B.S. degree from the Department of Electrical Engineering, National Chung Cheng University, Chiayi, Taiwan, in 2023. 
He is currently pursuing the Ph.D. degree with the Intelligent Hyperspectral Computing Laboratory, Institute of Computer and Communication Engineering, National Cheng Kung University (NCKU), Tainan, Taiwan. 
His research interests include deep learning, quantum weather forecasting, convex optimization, hyperspectral super-resolution, as well as interpretable AI. 
\end{IEEEbiography}

\begin{IEEEbiography}[{\resizebox{0.9in}{!}{\includegraphics[width=1in,height=1.25in,clip,keepaspectratio]{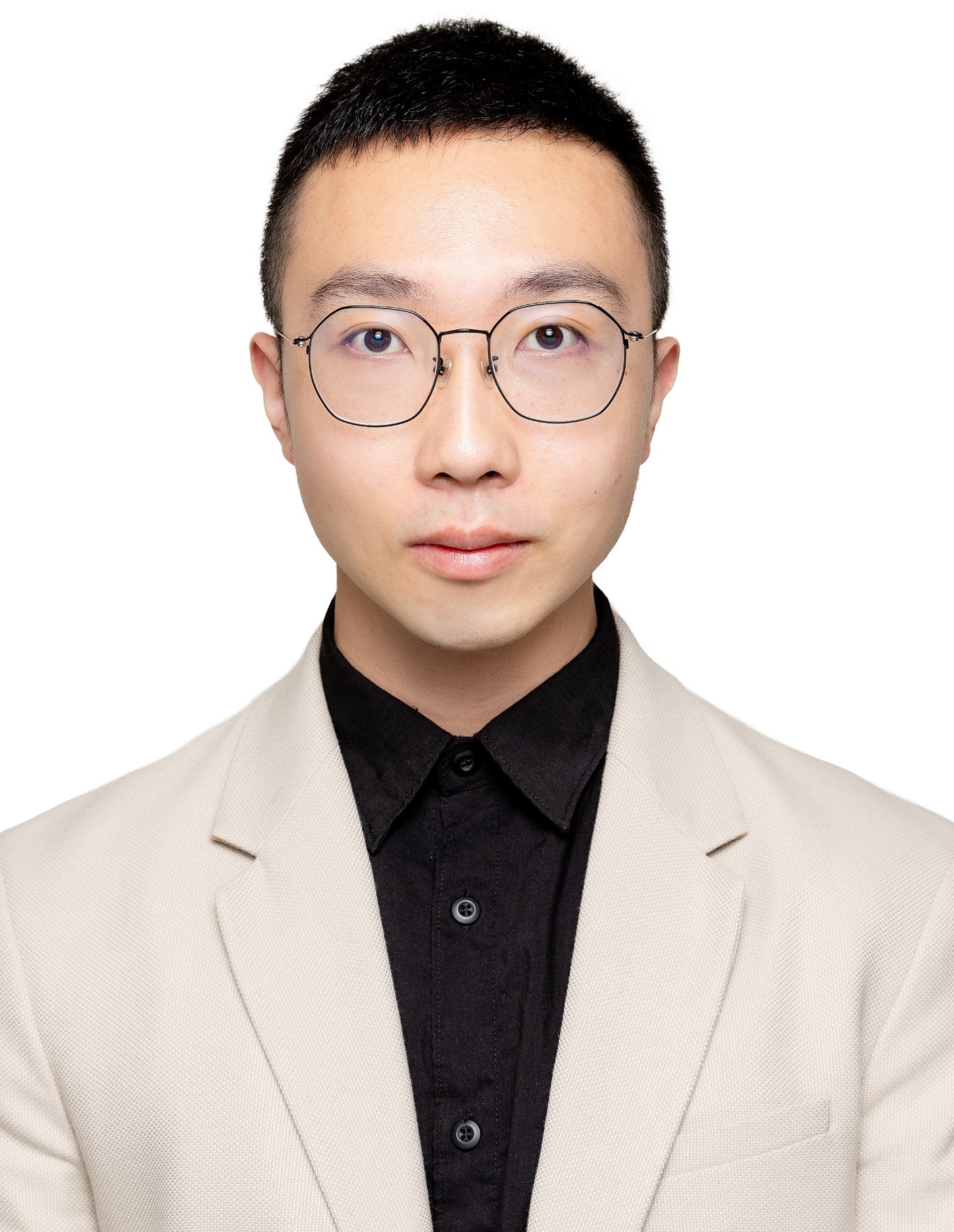}}}]
	{\bf Si-Sheng Young}
(S'23) is currently a Ph.D. student with the Intelligent
Hyperspectral Computing Laboratory (IHCL), Department of Electrical Engineering, National Cheng Kung University (NCKU), Tainan, Taiwan. 

In 2023, he received the Merit Award from The Grand Challenge ``Computing for the Future", Miin Wu School of Computing, NCKU, as well as the highly competitive ``Pan Wen Yuan Scholarship" from the Industrial Technology Research Institute (ITRI), Hsinchu, Taiwan.
In 2024, he received a highly competitive ``Scholarship Pilot Program to Cultivate Outstanding Doctoral Students" from the National Science and Technology Council (NSTC), Taiwan.
His research interests include convex optimization, deep learning, anomaly detection, data fusion, and imaging inverse problems.
\end{IEEEbiography}

\begin{IEEEbiography}[{\resizebox{1in}{!}{\includegraphics[width=1in,height=1.15in,clip,keepaspectratio]{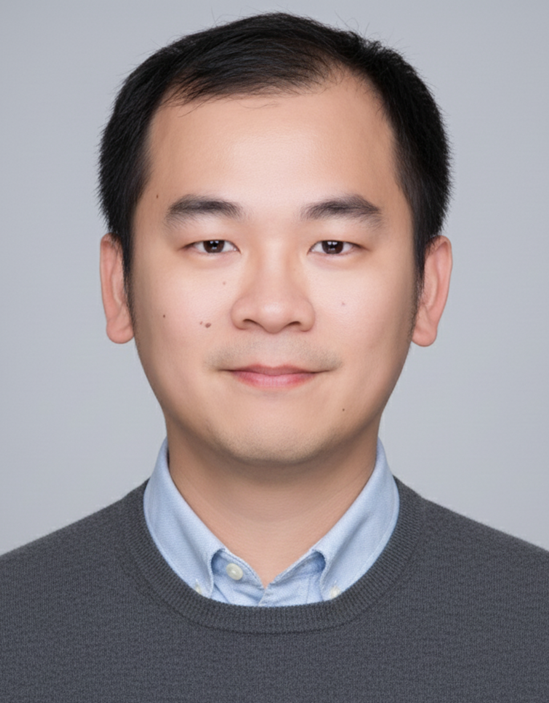}}}]
{\bf Wei-Cheng Zheng} received his B.S. degree from the Department of Geography, Chinese Culture University, Taiwan, in 2012, and his M.S. degree in Remote Sensing Science and Technology from the Center for Space and Remote Sensing Research (CSRSR), National Central University (NCU), Taiwan, in 2020. 
He is currently an Associate Research Fellow with the Agency of Rural Development and Soil and Water Conservation (ARDSWC), Ministry of Agriculture, Taiwan, since 2024, focusing on WebGIS-related projects. 
Prior to his current role, he was a Full-Time Research Fellow with CSRSR, NCU, for over nine years (2013–2024), where he specialized in using remote sensing theory, GIS software, and automation tools to support national-level projects, such as the ``Integrated Land Use Monitoring Operation''. 
%
His research interests include satellite remote sensing, land cover change detection, slopeland disaster prevention, and debris flow (landslide) investigation. 
\end{IEEEbiography}

\end{document}